\documentclass[12pt]{article}
\usepackage{epsfig,amsfonts,amsthm}
\usepackage{amsmath,amssymb}
\usepackage{cite}
\usepackage{srcltx}
\usepackage{tikz}
\newcommand{\be}{\begin{equation}}
\newcommand{\ee}{\end{equation}}
\newcommand{\bea}{\begin{eqnarray}}
\newcommand{\eea}{\end{eqnarray}}

\renewcommand{\Re}{\mathrm{Re }}
\renewcommand{\Im}{\mathrm{Im }}
\newcommand{\stolb}[3]{ \left( \begin{array}{c}#1 \\ #2 \\ #3\end{array}\right) }

\newcommand{\mmatrix}[4]{\left(\!\!\begin{array}{cc}#1 & #2 \\ #3 & #4 \end{array}\!\!\right)}

\newcommand{\Z}{\mathbb{Z}}
\newcommand{\F}{\mathbb{F}}
\newcommand{\arbitraryext}{\,\ldotp}

\providecommand{\CC}{{\mathbb{C}}}

\def\lsim{\mathrel{\rlap{\lower4pt\hbox{\hskip1pt$\sim$}}
    \raise1pt\hbox{$<$}}}         
\def\gsim{\mathrel{\rlap{\lower4pt\hbox{\hskip1pt$\sim$}}
    \raise1pt\hbox{$>$}}}         

\title{Classification of finite reparametrization symmetry groups in the three-Higgs-doublet model}

\author{Igor~P.~Ivanov$^{1,2}$, Evgeny Vdovin$^{2}$
\\
  {\small $^1$ IFPA, Universit\'{e} de Li\`{e}ge, All\'{e}e du 6 Ao\^{u}t 17, b\^{a}timent B5a, 4000 Li\`{e}ge, Belgium}\\
  {\small $^2$ Sobolev Institute of Mathematics, Koptyug avenue 4, 630090, Novosibirsk, Russia}\\
  }

\begin{document}
\maketitle

\begin{abstract}
Symmetries play a crucial role in electroweak symmetry breaking models with non-minimal Higgs content.
Within each class of these models, it is desirable to know which symmetry groups can be implemented via the scalar sector.
In $N$-Higgs-doublet models, this classification problem was solved only for $N=2$ doublets.
Very recently, we suggested a method to classify all realizable finite symmetry groups
of Higgs-family transformations in the three-Higgs-doublet model (3HDM).
Here, we present this classification in all detail together with an introduction
to the theory of solvable groups, which play the key role in our derivation.
We also consider generalized-$CP$ symmetries,
and discuss the interplay between Higgs-family symmetries and $CP$-conservation.
In particular, we prove that presence of the $\Z_4$ symmetry guarantees
the explicit $CP$-conservation of the potential.
This work completes classification of finite reparametrization symmetry groups in 3HDM.
\end{abstract}

\section{Introduction}

The nature of the electroweak symmetry breaking is one of the main puzzles
in high-energy physics.
Very recently, the CMS and ATLAS collaborations at the LHC
announced the discovery of the Higgs-like resonance at 126 GeV, \cite{discovery},
and their first measurements indicate intriguing
deviations from the Standard Model (SM) expectations.
Whether these data signal that a non-minimal Higgs mechanism is indeed at work 
and if so what it is, are among the hottest questions in particle physics these days.

In the past decades, many non-minimal Higgs sectors have been considered, \cite{CPNSh}.
One conceptually simple and phenomenologically attractive class of models involves
several Higgs doublets with identical quantum numbers ($N$-Higgs-doublet models, NHDM).
Its simplest version with only two doublets, 2HDM, was proposed decades ago, \cite{TDLee},
but it is still actively studied, see \cite{review2011} for a recent review,
and it has now become a standard reference model of the beyond the Standard Model (bSM) physics.
Constructions with more than two doublets are also extensively investigated,
\cite{weinberg3HDM,variants,A43HDM,toorop,branco-gerard-grimus84,varzielas,nishiNHDM2007,FerreiraSilva2008,ivanovNHDM,ivanovNHDM2}.

Many bSM models aim at providing a natural explanation for the numerical values of (some of) the SM parameters.
Often, it is done by invoking additional symmetries in the model.
These are not related with the gauge symmetries
of the SM but rather reflect extra symmetry structures in the ``horizontal space'' of the model.
One of the main phenomenological motivations in working with several doublets is the ease
with which one can introduce various symmetry groups. 
Indeed, Higgs fields with identical quantum numbers can mix, 
and it is possible that some of these Higgs-family mixing transformations leave the scalar sector invariant.
Even in 2HDM, presence of such a symmetry in the lagrangian and its possible spontaneous violation
can lead to a number of remarkable phenomena such as various forms of $CP$-violation, \cite{TDLee,CPunusual},
non-standard thermal phase transitions which may be relevant for the early Universe, \cite{thermal},
natural scalar dark matter candidates, \cite{IDM}.
For models with three or more doublets, an extra motivation is the possibility to incorporate
into the Higgs sector non-abelian finite symmetry groups, which can then lead to interesting patterns
in the fermionic mass matrices (for a general introduction into discrete symmetry groups relevant for
particle physics, see \cite{discrete-particle}). In this respect, the very popular symmetry group
has been $A_4$, \cite{A43HDM,toorop}, the smallest finite group with
a three-dimensional irreducible representation, but larger symmetry groups also received some attention,
\cite{weinberg3HDM,branco-gerard-grimus84,varzielas}.

Given the importance of symmetries for the NHDM phenomenology,
it is natural to ask: {\em which symmetry groups can be implemented in the scalar sector of NHDM
for a given $N$?}

In the two-Higgs-doublet model (2HDM), this question has been answered several years ago,
\cite{2HDMsymmetries,ivanov2HDM}, see also \cite{review2011} for a review.
Focusing on discrete symmetries, the only realizable Higgs-family symmetry groups
are $\Z_2$ and $(\Z_2)^2$. The $\Z_2$ group can be generated, for example,
by the sign flip of one of the doublets (and it does not matter which, because
once we focus on the scalar sector only, the simultaneous sign flip of both doublets does not change the lagrangian),
while the $(\Z_2)^2$ group is generated by sign flips and the exchange $\phi_1 \leftrightarrow \phi_2$.
If generalized-$CP$ transformations are also included, then $(\Z_2)^3$ becomes realizable as well,
the additional generator being simply the $CP$ conjugation.

With more than two doublets, the problem remains open.
Although several attempts have been made in past to classify at least some symmetries in NHDM,
\cite{nishiNHDM2007,FerreiraSilva2008,ivanovNHDM2}, they led only to very partial results.
The main obstacle here was the lack of the completeness criterion.
Although many obvious symmetry groups could be immediately guessed,
it was not clear how to prove that the given potential does not have other symmetries.
An even more difficult problem is to prove that no other symmetry group can be implemented for a given $N$.

In the recent paper \cite{abelianNHDM} we found such a criterion
for abelian symmetry groups in NHDM for arbitrary $N$.
Since abelian subgroups are the basic building blocks of any group,
classification of realizable abelian symmetry groups in NHDM was an important milestone.
We stress that this task is different from just classifying all abelian subgroups of $SU(3)$,
because invariance of the Higgs potential places strong and non-trivial restrictions on possible symmetry groups.

In this paper, we solve the classification problem for all finite symmetry groups in 3HDM, including non-abelian groups.
We do this by using the abelian groups in 3HDM found in \cite{abelianNHDM}
and by applying certain results and methods from the theory of solvable groups.
Some of these results were already briefly described in \cite{finite3HDMshort}.
Here, we present a detailed derivation of this classification together with an introduction to
the relevant methods from finite group theory. In addition, we extend
the analysis to symmetry groups which include both Higgs-family and generalized-$CP$ transformations.
This work, therefore, solves the problem of classification of finite reparametrization
symmetry groups in 3HDM.

We would like to stress one important feature in which our method differs from more traditional approaches to symmetry 
classification problem, at least within the bSM physics. 
Usually, one starts by imposing invariance under certain transformations,
and then one tries to recognize the symmetry group of the resulting potential.
In this way it is very difficult to see whether all possible symmetries are exhausted.
We approach the problem the other way around.
We first restrict the list of finite groups which can appear as symmetry groups of 3HDM,
and then we check one by one whether these groups can indeed be implemented.
\\

The structure of this paper is the following. In Section~\ref{section-symmetries}
we describe different types of symmetries in the scalar sector of NHDM and discuss the important
concept of realizable symmetry groups.
Section~\ref{section-solvable} contains an elementary introduction into the theory of (finite) solvable groups.
Although it contains pure mathematics, we put it in the main text because it is a key part of the group-theoretic step of our classification, 
which is presented in Section~\ref{section-structure}.
Then, in Section~\ref{section-continuous} we describe the methods which we will use to prove the absence of
continuous symmetries.
Sections~\ref{section-torus} and \ref{sectionZ3Z3} contain the main results of the paper: explicit constructions
of the realizable symmetry groups and of the potentials symmetric under each group.
Finally, in Section~\ref{section-summary} we summarize and discuss our results.
For the reader's convenience, we list in the Appendix potentials for each of the realizable 
non-abelian symmetry groups.

\section{Symmetries of the scalar sector of multi-Higgs-doublet models}\label{section-symmetries}

\subsection{Reparametrization transformations}

In NHDM we introduce $N$ complex Higgs doublets with the electroweak isospin $Y=1/2$,
which interact with the gauge bosons and matter fields in the standard way,
and also self-interact via a Higgs potential.
The generic renormalizable Higgs potential can contain
only quadratic and quartic gauge-invariant terms, and it can be compactly written as \cite{tensorial-form,CPviolation}:
\be
\label{V:tensorial}
V = Y_{ab}(\phi^\dagger_a \phi_b) + Z_{abcd}(\phi^\dagger_a \phi_b)(\phi^\dagger_c \phi_d)\,,
\ee
where all indices run from 1 to $N$.
Coefficients of the potential are grouped into components of tensors $Y_{ab}$ and $Z_{abcd}$;
there are $N^2$ independent components in $Y$ and $N^2(N^2+1)/2$ independent components in $Z$.

In this work we focus only on the scalar sector of the NHDM.
Therefore, once coefficients $Y_{ab}$ and $Z_{abcd}$ are given, the model is completely defined,
and one should be able to express all its properties (the number and the positions of extrema, the spectrum
and interactions of the physical Higgs bosons) via components of $Y$'s and $Z$'s.
This explicit expression, however, cannot be written via elementary functions, 
and it remains unknown in the general case for any $N > 2$.

A very important feature of the most general potential is that any non-degenerate linear
transformation in the space of Higgs doublets belonging to the group $GL(2,\CC)$ keeps the generic form of the potential,
changing only the coefficients of $Y$ and $Z$. We call such a transformation
a {\em Higgs-basis change}.
In addition, the $CP$ transformation, which maps doublets to their hermitean conjugates $\phi_a \to \phi_a^\dagger$,
also keeps the generic form of the potential, up to coefficient modification.
Its combination with a Higgs-basis change represents a transformation
which is usually called a {\em generalized-$CP$ transformation}, \cite{GCP}.
The Higgs basis changes and generalized-$CP$ transformations can be called together
{\em reparametrization transformations} because they preserve the generic structure of the potential
and lead only to its reparametrization.

A reparametrization transformation changes the basis in the space of Higgs doublets but does not modify the structural features
of the model such as the number and the properties of minima, the symmetries of the potential and their spontaneous breaking
at the minimum point.
These properties must be the same for all the potentials linked by reparametrization transformations.
Therefore, these properties must be expressible in terms of {\em reparametrization-invariant} combinations of
$Y$'s and $Z$'s, \cite{tensorial-form,haber}.

If a reparametrization transformation maps a certain potential exactly to itself, that is, 
if it leaves certain $Y$'s and $Z$'s invariant,
we say that the potential has a {\em reparametrization symmetry}.
Usually, there is a close relation between the reparametrization symmetry group $G$ of the potential
and its phenomenological properties, both within the scalar and the fermion sectors.
Therefore, understanding which groups can appear as reparametrization symmetry groups in NHDM with given $N$
is of much importance for phenomenology of the model.

\subsection{The group of kinetic-term-preserving reparametrization transformations}

Often, one restricts the group of reparametrization transformations only to those transformations
which keep the Higgs kinetic term invariant.
In this case, a generic basis change becomes a unitary transformation $\phi_a \mapsto U_{ab}\phi_b$
with $U\in U(N)$.
A kinetic-term-preserving generalized-$CP$ transformation is an anti-unitary map $\phi_a \mapsto U_{ab}\phi^\dagger_b$,
which can be written as $U_{CP} = U \cdot J$, with a unitary $U$ and with $J$ being the symbol for the $CP$-transformation.

The group $U(N)$ contains the group of overall phase rotations, which are already included in the gauge group $U(1)_Y$.
Since we want to study structural symmetries of the NHDM potentials, we should disregard transformations
which leave all the potentials invariant by construction. This leads us to the group $U(N)/U(1) \simeq PSU(N)$.
Note that $SU(N)$, which is often considered in these circumstances, 
still contains transformations which only amount to the overall phase shift of all doublets.
They form the center of $SU(N)$, $Z(SU(N)) \simeq \Z_N$, and act trivially on all NHDM potentials.
Being invariant under them does not represent any structural property of the Higgs potential,
therefore, we are led again to the factor group $SU(N)/Z(SU(N)) = PSU(N)$.
This allows us to write the group of kinetic-term-preserving reparametrization transformations as a semidirect product
of the Higgs basis change group and the $\Z_2$ group generated by $J$ (for a more detailed discussion, see \cite{abelianNHDM}):
\be
G_{rep} = PSU(N) \rtimes \Z_2^*\,.\label{Grep}
\ee
Here the asterisk indicates that the generator of the corresponding group is an anti-unitary transformation;
we will use this notation throughout the paper.

Below, when discussing symmetry groups of the 3HDM potential, we will be either looking for subgroups of $PSU(3)$
(if only unitary transformations are allowed) or subgroups of this $G_{rep}$ (when anti-unitary reparametrization transformations
are also included). This should always be kept in mind when comparing our results with the groups which are
discussed as symmetry groups in the 3HDM scalar sector. For example, in \cite{branco-gerard-grimus84,varzielas}
a 3HDM potential symmetric under $\Delta(27)$ or $\Delta(54)$ was considered, both groups being subgroups
of $SU(3)$. However, they contains the center of $SU(3)$, which, we repeat, acts trivially on all Higgs potentials.
Therefore, the structural properties of that model are defined by the factor groups $\Delta(27)/Z(SU(3)) \simeq \Z_3\times\Z_3$
and $\Delta(54)/Z(SU(3)) \simeq (\Z_3\times\Z_3)\rtimes \Z_2$, which belong to $PSU(3)$.

\subsection{Realizable symmetry groups}\label{subsection-realizable}

There is an important technical point which should be kept in mind when we classify symmetry groups of NHDM.
When we impose a reparametrization symmetry group $G$ on the potential, we restrict its coefficients in a certain way.
It might happen then that the resulting potential becomes symmetric under a {\em larger} symmetry group $\widetilde{G}$
properly containing $G$.

One drawback of this situation is that we do not have control over the true symmetry properties of the potential:
if we construct a $G$-symmetric potential, we do not know a priori what is its full symmetry group $\widetilde{G}$.
This might be especially dangerous if $G$ is finite while $\widetilde{G}$ turns out to be continuous, as it might lead
to unwanted goldstone bosons.
Another undesirable feature is related with symmetry breaking.
Suppose that we impose invariance of the potential under group $G$ but we do not check what is
the true symmetry group $\widetilde{G}$.
After electroweak symmetry breaking, the symmetry group of the vacuum is $G_{v} \leq \widetilde{G}$,
and it can happen that $G_v$ is {\em not} a subgroup of $G$.
This is not what we normally expect when we construct a $G$-symmetric model, and it is an indication of a higher symmetry.

Examples of these situations were encountered in literature before. For instance,
the authors of \cite{FerreiraSilva2008} explicitly show that trying to impose a $\Z_p$, $p > 2$, group of rephasing transformations
in 2HDM unavoidably leads to a potential with continuous Peccei-Quinn symmetry.
For 3HDM they find an even worse example, when a cyclic group immediately leads to a $U(1)\times U(1)$-symmetric potential.
Another well-known example is the $A_4$-symmetric 3HDM potential, which at certain values of parameters
admits vacua with the $S_3$ symmetry, although $S_3$ is not a subgroup of $A_4$, see an explicit study in \cite{toorop}. The explanation is that
the potential at these values of parameters becomes symmetric under $S_4$
which contains both $A_4$ and $S_3$.

In order to avoid such situations altogether, we must always check
for each $G$ whether the $G$-symmetric potentials are invariant under any larger group.
We are interested only in those groups $G$, for which there exists a $G$-invariant potential with the property 
that no other reparametrization transformation leaves it invariant (either within $PSU(3)$ or within $G_{rep}$,
depending on whether we include anti-unitary transformations). Following \cite{ivanovNHDM2,abelianNHDM},
we call such groups {\em realizable}.

Using the terminology just introduced we can precisely formulate the two main questions which we address in this paper:
\begin{enumerate}
\item
considering only non-trivial kinetic-term-preserving Higgs-basis transformations (i.e. group $PSU(3)$),
what are the realizable finite symmetry groups in 3HDM?
\item
more generally, considering non-trivial kinetic-term-preserving reparametrization transformations, which can now include
generalized-$CP$ transformations (i.e. group $G_{rep}$), what are the realizable finite symmetry groups in 3HDM?
\end{enumerate}
For abelian groups, these questions were answered in \cite{abelianNHDM} for general $N$.
Here we focus on non-abelian finite realizable groups for $N=3$.

\section{Solvable groups: an elementary introduction}\label{section-solvable}

Our classification of realizable groups of Higgs-family symmetries in 3HDM
contains two essential parts: the group-theoretic and the calculational ones.
The group-theoretic part will make use of some methods of pure finite group theory,
which are not very familiar to the physics community
(although they are quite elementary for a mathematician with expertise in group theory).
To equip the reader with all the methods needed to understand
the group-theoretic part of our analysis,
we begin by giving a concise introduction to the theory of solvable groups.
In doing so, we mention only methods and results which are relevant for the particular problem of this paper.
For a deeper introduction to solvable groups and finite group theory in general,
see e.g. \cite{Isaacs}.

\subsection{Basics}

We assume that the reader is familiar with the basic definitions from group theory.
We only stress here that we will work with finite groups, therefore the {\em order} of the group $G$ 
(the number of elements in $G$) denoted as $|G|$
is always finite, and so is the order of any element $g$ (the smallest positive integer $n$ such that $g^n = e$, 
the identity element of the group).

A group $G$ is called abelian if all its elements commute. An alternative way to formulate it is to say
that all {\em commutators} in the group are trivial: $[x,y] = xyx^{-1}y^{-1}=e$ for all $x,y \in G$.
Working with commutators is sometimes easier than checking the commutativity explicitly. For example,
it is easy to prove that if every non-trivial element of the group has order two, $g^2 = e$, then the group is abelian.
Indeed, for any $x,y \in G$ we have
\be
[x,y]=xyx^{-1}y^{-1}=xyxy=(xy)^2=e\,,
\label{every-element-has-order-2}
\ee
which means that $x$ and $y$ commute.

A group $G$ can have proper subgroups $H  <  G$ (whenever we do not require
that the subgroup $H$ is proper, we write $ H \leq G$), whose order must, by Lagrange's theorem,
divide the order of the group: $|H|$ divides $|G|$. If proper subgroups exist, some of them must be abelian.
A simple way to obtain an abelian subgroup is to pick up an element $g \in G$ and consider its powers:
if order of the element $g$ is $n$, we will get the cyclic group $\Z_n  <  G$.

The inverse of Largrange's theorem is not, generally speaking, true: namely, if $p$ is a divisor of $|G|$,
the group $G$ does not necessarily have a subgroup of order $p$.
However, if $p$ is a prime which enters the prime decomposition of $|G|$, then according to Cauchy's theorem
such a subgroup must exist (this group is $\Z_p$ because there are no other groups of prime order).
It immediately follows that if we have the list of all abelian subgroups of a given finite group $G$,
then the prime decomposition of $|G|$ can only contain primes which are present in the orders 
of these abelian subgroups.

In fact, there is an existence criterion stronger than Cauchy's theorem.
Namely, if $p^a$ is the highest power of the prime $p$ that enters
the prime decomposition of $|G|$, then $G$ contains a subgroup of this order,
which is called the Sylow $p$-subgroup of the group $G$. This theorem (known as the Sylow-E theorem)
is the starting point of the theory of Sylow subgroups, see Chapter 1 in \cite{Isaacs}.

There are several ways to {\em present} a finite group. One possibility is to list all its elements and write down
the $|G|\times |G|$ multiplication table. Clearly, this presentation becomes impractical for
a sufficiently large group. A more compact and powerful way is known as {\em presentation by generators
and relations}. We call a subset $M =\{g_1,\, g_2,\,\dots\}$ of the elements of $G$ a {\em generating set}
(and its elements are called {\em generators})
if every $g \in G$ can be written as a product of elements of $M$ or their inverses.
The fact that $G$ is generated by the set $M$ is denoted as $G = \langle M\rangle$.
Finding a minimal generating set for a given group and listing equalities which these generators satisfy
is precisely presentation of the group by generators and relations.
For example, the symmetry group of the regular $n$-sided polygon
has the following presentation by generators and relations:
\be
D_{2n} = \langle a,b\, | \,a^2=b^2=(ab)^n = e\rangle\,.
\ee
This group is known as the {\em dihedral group} and has order $|D_{2n}|=2n$ 
(note that there exists an alternative convention for denoting dihedral groups: $D_n$;
the one which we use has its order in the subscript).

\subsection{Normal subgroups and extensions}

Consider two groups $G$ and $H$. Suppose we have a map $f$ from $G$ to $H$, $f: G \to H$,
which sends every $g\in G$ into its image $f(g) \in H$.
If this map preserves the group operation, $f(g_1)f(g_2)=f(g_1g_2)$,
then it is called a {\em homomorphism}. If this map is surjective (i.e. it covers the entire $H$)
and injective (distinct elements from $G$ have distinct images in $H$), then $f$ is invertible and
is called an isomorphism.

In the case when $H=G$, we deal with an isomorphism of the group onto itself,
which is called an {\em automorphism}.
One can note that composition of two automorphisms is also an automorphism,
and define the group structure on the set of all automorphisms of $G$.
This {\em automorphism group} is denoted as $Aut(G)$.
The trivial automorphism which fixes every element of $G$ is the identity element of~$Aut(G)$.

Let us now consider a special class of automorphisms called {\em inner} automorphisms, or {\em conjugations}.
Fix an element $g\in G$ and define $f: x \mapsto g^{-1}xg$ for every $x \in G$.
It can be immediately checked that   $f$ is an automorphism, and that it sends a subgroup of $G$ into a
(possibly another) subgroup of $G$. It can however happen that certain subgroups will be mapped onto themselves:
$g^{-1}Hg = H$. Subgroups which satisfy this invariance criterion for every possible $g \in G$ are called {\em normal},
or invariant subgroups. The fact that $H$ is a normal subgroup of $G$ is denoted as $H \lhd G$.

Even when a subgroup $H$ is not normal in $G$, one can pick up some elements $g\in G$ such that $g^{-1}Hg=H$.
The set of elements of  $G$ with the property $g^{-1}Hg=H$ forms a group, which is called 
the {\em normalizer} of $H$ in $G$ and denoted as $N_G(H)$.
We then have $H \lhd N_G(H) \leq G$. Working with normalizers is a useful intermediate step in situations
when it is not known whether the subgroup $H$ is normal in $G$.

Having a normal subgroup $H \lhd G$ gives some information about the structure of $G$.
One can define the group structure on the set of (left) cosets of $H$,
which is now called the {\em factor group} $G/H$. Thus, one breaks the group
into two smaller groups, which often simplifies its study.
Given a normal subgroup $H \lhd G$, one can define the {\em canonical homomorphism}
$\phi: G \to G/H$ which sends every element $g\in G$ into its coset $gH$. Its kernel
(all elements $g$ which are mapped by $\phi$ into the identity element of $G/H$) is precisely $H$.
Thus, every normal subgroup is the kernel of the corresponding canonical homomorphism.
The reverse statement is also true: kernels of homomorphisms are always normal subgroups.

The group-constructing procedure inverse to factoring is called {\em extension}.
Given two groups, $N$ and $H$, a group $G$ is called an extension of $H$ by $N$ (denoted as $N\arbitraryext H$),
if there exists $N_0 \lhd G$ such that $N_0 \simeq N$ and $G/N_0 \simeq H$.
In the case when, in addition, $H$ is also isomorphic to a subgroup of $G$ and $G=NH$,
we deal with a {\em split extension}. The criterion for $G$ to be a split extension
can also be written as existence of $N \lhd G$ and $H \leq G$ such that $NH = G$ and $N\cap H = 1$,
so that $G/N = H$.
The group $G$ is then called a {\em semidirect product} $G = N \rtimes H$.

Even if two groups $N$ and $H$ are fixed, they can support several extensions and split extensions.
Therefore one faces the problem of classifying of all extensions of two given groups.

For the most elementary example, consider extensions of $H=\Z_2$ (generated by $a$) by $N=\Z_2$ (generated by $b$),
which should produce a group of order 4.
Then, for a split extension, we need a group $G$ which has two distinct subgroups
isomorphic to $N$ and $H$. The only choice is $G = \Z_2\times \Z_2$, which can be presented as $\langle a,b\, |\,
a^2=b^2=(ab)^2=e\rangle$.
For a non-split extension, we require that only $N$ is isomorphic to a subgroup of $G$.
Thus, we still have $b^2 = e$, while $a^2$ must {\em not} be the unit element.
Then we have to set $a^2 = b$ producing the group $\Z_4$.
So, $\Z_4$ does not split over $\Z_2$, while $\Z_2\times \Z_2$ does.

\subsection{Characteristic subgroups}

In what concerns embedding of groups, normality is a relatively weak property.
Namely, if $K \lhd H$ and $H \lhd G$, then $K$ is not necessarily normal in $G$
(it is instead called subnormal in $G$).
Indeed, recall that a normal subgroup $K \lhd H$ stays invariant under all inner automorphisms on $H$.
Here ``inner'' is meant with respect to the group $H$, namely, $h^{-1}Kh=K$ for all $h \in H$.
However since $H \lhd G$, one can fix $g \in G$ but $g \not \in H$ and consider
an automorphism on $H$ defined by $H \to g^{-1}Hg$. This is indeed an automorphism
on $H$ because it induces a permutation of elements of $H$ preserving its group property,
but it is not inner, because $g$ does not belong to $H$.
Therefore $K$ does not have to be invariant under it: $g^{-1}Kg \not = K$.

However there is a stronger property which guarantees normality for embedded groups.
Let us call a subgroup $K$ {\em characteristic} in $H$ if it is invariant under all (not only inner) automorphisms
of $H$. Then, repeating the above arguments, we see that if $K$ is characteristic in $H$, and $H$
is normal in $G$, then $K$ is also normal in $G$. Also, if $K$ is characteristic in $H$
and $H$ is characteristic in $G$, then $K$ is also characteristic in $G$.
Thus, knowing that some subgroups
are characteristic gives even more information than their normality.

There is one simple rule which guarantees that certain subgroups are characteristic.
If we have a rule defined in terms of the group $G$ which identifies its subgroup $H$ uniquely, then $H$ is characteristic in $G$.
Two important examples are:
\begin{itemize}
\item
the {\em center} of the group $G$ denoted as $Z(G)$, which is the set of all elements $z \in G$ such that they commute
with all elements of $G$:
\be
Z(G) = \{z\in G\, |\, [z,g]=e\ \forall g \in G\}\,.\label{center}
\ee
The center of an abelian group coincides with the group itself.
\item
the {\em commutator subgroup} (or {\em derived subgroup}) of $G$
denoted as $G'$ and defined as the subgroup generated by all commutators:
\be
G' = \langle [x,y] \rangle\,,\quad x,y \in G\,. \label{commutator-group}
\ee
Note that the word ``generated'' is needed because the set of commutators is generally speaking
not closed under the group multiplication. Clearly, the commutator subgroup of an abelian group is trivial,
therefore the size of $G'$ can be used to qualitatively characterize how far $G$ is from being abelian.
\end{itemize}

\subsection{Consequences of existence of a normal maximal abelian subgroup}\label{subsection-maximal-normal-abelian}

Let us now prove a rather simple group-theoretic result, which however will be important
for our classification of symmetries in 3HDM.
This result, loosely speaking, is the observation that a mere existence of a subgroup of $G$
with some special properties can strongly restrict the structure of the group $G$.

First, an abelian subgroup $A  <  G$ is called  a {\em maximal abelian subgroup} if
there is no other abelian subgroup $B$ with property $A  <  B \leq G$.
Note that the word ``maximal'' refers not to the size but to containment.
This definition does not specify a unique subgroup;
in fact a group can have several maximal abelian subgroups.
They correspond to terminal points in the partially-ordered tree of abelian subgroups of $G$.

Suppose that $A$ is an abelian subgroup of a finite group $G$.
Elements of $A$, of course, commute among themselves. But it can also happen that
there exist other elements $g \in G$, $g \not \in A$, which also commute with all elements of $A$.
The set of all such elements is called the {\em centralizer} of $A$ in $G$:
\be
C_G(A) = \{g \in G\,|\, [g,a]=e\ \forall a \in A\}\,.
\ee
It is easy to check that $C_G(A)$ is a subgroup of $G$, and it can be non-abelian.
The name ``centralizer'' refers to the fact that although $A$ is not the center in $G$, it is the center in $C_G(A)$.

Clearly, $A \leq C_G(A)$. If $A$ is a proper subgroup of $C_G(A)$, then it means that $A$ is not a maximal abelian subgroup.
Indeed, we take an element $g \in C_G(A)$, $g \not \in A$, and consider another subgroup
$B = \langle A, g\rangle$. This subgroup is abelian and is strictly larger than $A$: $A  <  B \leq G$.
On the other hand, an element $x \in G$ which commutes with all elements of $B$
will certainly commute with all elements of $A$, while the converse is not necessarily true.
Therefore, we get the following chain:
$A  <  B \leq C_G(B) \leq C_G(A)$.
Next, we check whether $B$ is a proper subgroup of $C_G(B)$. If so, we can enlarge it again in the same way
by considering $C = \langle B, g'\rangle$, where $g' \in C_G(B)$, $g' \not \in B$.
We can continue this procedure until it terminates with an abelian subgroup $K$ which is {\em self-centralizing}:
\be
A  <  B  <  \cdots  <  K = C_G(K) \leq \cdots \leq C_G(B) \leq C_G(A)\,.
\ee
Since there exists no other element in $G$ which would commute with all elements of $K$,
we conclude that $K$ is a maximal abelian subgroup in $G$.

Let us now see what changes if the abelian subgroup $A$ is normal. Any element $g \in G$ acting on $A$ by conjugation
induces an automorphism of $A$. Thus, we have a map from $G$ to the group of automorphisms of $A$,
$f: G \to Aut(A)$. The kernel of $f$ consists of such $g$'s which induce the trivial automorphism of $A$,
that is, which leave every $a \in A$ unchanged: $g^{-1}ag=a$ $\forall a \in A$. But this coincides with the definition of centralizer.
Therefore we conclude that $\ker f = C_G(A)$.

The fact that $C_G(A)$ is the kernel of the homomorphism $f$ implies that $C_G(A)$ is a normal subgroup of $G$.
Note that it is essential that the abelian subgroup in question, $A$, is normal; if it were not, $C_G(A)$  would not have to be normal.

Now, if $A$ is a normal maximal abelian subgroup of $G$, then $\ker f = C_G(A) = A$. In other words,
the kernel of $G/A \to Aut(A)$ is trivial, and therefore, $G/A$ is isomorphic to a subgroup of $Aut(A)$.
Summarizing our discussion, if $A$ is a normal maximal abelian subgroup of $G$, then $G$ can be constructed
as an extension of $A$ by a subgroup of $Aut(A)$:
\be
G \simeq A\arbitraryext K\,,\quad \mbox{where}\quad K \leq Aut(A)\,.\label{structure-G}
\ee
This is a powerful structural implication for the group $G$ of existence of a normal maximal abelian subgroup.

\subsection{Automorphism groups}

For future reference, we give some details on the automorphism groups $Aut(A)$ of certain abelian groups $A$.
In this subsection we will use the additive notation for the group operation.

Suppose $A=\Z_n$ is the cyclic group of order $n$ with generator $e$:
$n e = \underbrace{ e + \dots + e }_{n\text{ times}}= 0$.
An automorphism $\sigma$ acting on $A$ is a group-structure-preserving permutation of elements of $A$.
Since $A$ is generated by $e$, this automorphism is completely and uniquely defined
once we assign the value of $\sigma(e)= k$ and make sure that
$m \sigma(e) \not = 0$ for all $0 < m < n$. This holds when $k$ and $n$ are coprime ($k=1$ is coprime to any $n$).
The number of integers less than $n$ and coprime to $n$
is called the Euler function $\varphi(n)$. Thus, we have $|Aut(\Z_n)|=\varphi(n)$.
For a prime $p$, the Euler function is obviously $\varphi(p) = p-1$. In general, if $p_1^{k_1}\cdots p_s^{k_s}$
is the prime decomposition for $n$, then
$$
\varphi(p_1^{k_1}\cdots p_s^{k_s}) = \varphi(p_1^{k_1})\cdots \varphi(p_s^{k_s}) =
(p_1^{k_1}-p_1^{k_1-1})\cdots (p_s^{k_s}-p_s^{k_s-1})\,.
$$

Suppose now that $p$ is prime and
$$
A=\underbrace{\Z_p\times\cdots\times \Z_p}_{n\text{ times}}=(\Z_p)^n\,.
$$
Then $G$ can be considered as an $n$-dimensional vector space over a finite field $\F_p$ of order $p$.
Vectors in this space can be written as
$$
x = k_1 e_1 + \dots + k_n e_n\,,
$$
where numbers $k_i \in \F_p$ and ``basis vectors'' $e_i $ are certain non-zero elements
of the $i$-th group $\Z_p$.
The group of all automorphisms on $(\Z_p)^n$ is then the general linear group in this space $GL_n(p)$.

Again, in order to define an automorphism $\sigma$ acting on $A$,
it is sufficient to assign where the basis vectors $e_i$ are sent by $\sigma$ and
to make sure that they stay linearly independent:
that is, if $m_1 \sigma(e_1) + \dots + m_n \sigma(e_n) = 0$, with $m_i \in \F_p$, then
all $m_i = 0$.  In order to calculate $|GL_n(p)|$, we just need to find
to how many different bases the initial basis $\{e_1,\dots,e_n\}$ can be mapped to.
The first vector, $e_1$, can be sent to $p^n-1$ vectors, the second vector, $e_2$, can be then sent
to $p^n-p$ vectors linearly independent with $\sigma(e_1)$, and so forth. The result is
\be
|GL_n(p)| = (p^n-1)(p^n-p)\cdots (p^n-p^{n-1}) = p^{{n(n-1) \over 2}}(p-1)(p^2-1)\cdots(p^n-1)\,.\label{orderGLnp}
\ee
In particular, $|Aut(\Z_p\times\Z_p)|=|GL_2(p)|=p(p-1)(p^2-1)$,
and the $p$-subgroup of $Aut(\Z_p\times\Z_p)$ can only be $\Z_p$.

\subsection{Nilpotent groups}\label{subsection-nilpotent}

In group theory, a powerful tool to investigate structure and properties of groups is
to establish existence of subgroup series with certain properties.
For example, a finite collection of normal subgroups $N_i \lhd G$ is called a {\em normal series} for $G$ if
\be
1 = N_0 \leq N_1 \leq N_2 \leq \cdots \leq N_r = G\,.\label{normal-series}
\ee
Restricting the properties of the factor groups $N_i/N_{i-1}$ for all $i$, one can infer non-trivial consequences
for the group $G$.

If all the factor groups in the normal series  lie in the centers, $N_i/N_{i-1} \leq Z(G/N_{i-1})$ for
$1 \leq i \leq r$, then (\ref{normal-series}) becomes a {\em central series}, and the group $G$ is then called
{\em nilpotent}. The smallest number $r$ for which the central series exists is called the nilpotency class of $G$.

Clearly, abelian groups are nilpotent groups of class 1 because for them $G \leq Z(G)$.
A non-abelian group $G$ whose factor group by its center $G/Z(G)$ gives an abelian group is a nilpotent group of class 2,
etc. So, nilpotent groups are often regarded as ``close relatives'' of abelian groups in the class of non-abelian ones.
One important class of nilpotent groups is $p$-groups, i.e. finite groups whose order is a power of a prime $p$.

Nilpotent groups bear several remarkable features. We mention here only two of them which we will use below.
First, a nilpotent group has a normal self-centralizing, and therefore maximal, abelian subgroup (Lemma 4.16 in \cite{Isaacs}),
whose implications were discussed above.
Second, if $H$ is a proper subgroup of a nilpotent group $G$, then $H$ is also a proper subgroup of $N_G(H)$ (Theorem 1.22 in \cite{Isaacs}).
In other words, the only subgroup of a nilpotent group $G$ which happens to be self-normalizing is the group $G$ itself.

\subsection{Solvable groups}

A group $G$ is called {\em solvable} if it has a normal series (\ref{normal-series})
in which all factor groups $N_i/N_{i-1}$ are abelian.
This is a broader definition than the one of nilpotent groups.
Therefore we can expect that both criteria and properties of solvable groups will be weaker than for nilpotent groups.

One particular example is that unlike nilpotent groups, a solvable group does not have to possess
a normal self-centralizing abelian subgroup. However what it does possess is just a {\em normal abelian subgroup}.
In order to prove this statement, let us first introduce another series of nested subgroups,
called the {\em derived series}. We first find $G'$, the derived subgroup of $G$, then we find its derived subgroup,
$G'' = (G')'$, then the third derived subgroup, $G^{(3)}=(G'')'$, and so on.
The derived series is simply
\be
\cdots \leq G^{(3)} \leq G'' \leq G' \leq G\,.\label{derived-series}
\ee

The relation of the derived series with solvability is the following: $G$ is solvable if and only if its derived series terminates,
i.e. $G^{(m)}=1$ for some integer $m\ge 0$ (Lemma 3.9 in \cite{Isaacs}).
The basic idea behind the proof of this statement is the observations that $G'$ is the unique smallest normal subgroup
of $G$ with an abelian factor group. Indeed, if $N\lhd G$ and $\phi: G \to G/N$ is the canonical homomorphism,
then $\phi(G')=(G/N)'$ (commutators are mapped into commutators).
If we want $G/N$ to be abelian, then $(G/N)' = 1$, and $G' \leq \ker \phi = N$.
Therefore, whatever $N_{r-1}$ we choose in (\ref{normal-series}), it will contain $G'$.
This argument can be continued through the series, and since the normal series terminates, so does the derived series.

Now, since $G^{(m)}=1$ for some finite $m$, we can consider $G^{(m-1)}$. It is an abelian group
because its derived subgroup is trivial. Being a characteristic subgroup of $G^{(m-2)}$, it is definitely normal in $G$.
Thus, we obtain the desired normal abelian subgroup.

A normal abelian subgroup is not guaranteed to be maximal.
One can, of course, extend it to a maximal abelian subgroup, but then it is not guaranteed to be normal.
Thus, in order to use the result (\ref{structure-G}), we need to prove the existence of an abelian subgroup
which combines both properties.
This situation is not generic: a solvable groups does not have to possess a normal maximal abelian subgroup.
However it can possess it in certain cases, and we will show below that in what concerns finite symmetry groups in 3HDM,
they do contain such a subgroup.

\section{Structure of the finite symmetry groups in 3HDM}\label{section-structure}

\subsection{Abelian subgroups and Burnside's theorem}

Our goal is to understand which finite groups $G$ can be realized as Higgs-family symmetry groups
in the scalar sector of 3HDM. We stress that we look for realizable groups only, see discussion
in section~\ref{subsection-realizable}.

Since finite groups have abelian subgroups, it is natural first to ask which abelian
subgroups $G$ can have. This can be immediately inferred from our paper \cite{abelianNHDM}
devoted to abelian symmetry groups in NHDM.
In the particular case of 3HDM, only the following groups can appear as abelian subgroups
of a finite realizable symmetry group $G$:
\be
\Z_2\,, \quad \Z_3\,, \quad \Z_4\,, \quad \Z_2\times \Z_2\,,\quad \Z_3 \times \Z_3\,.\label{abelian}
\ee
The first four are the only realizable finite subgroups of maximal tori in $PSU(3)$.
The last group, $\Z_3 \times \Z_3$, is on its own a maximal abelian subgroup of $PSU(3)$,
but it is not realizable because a $\Z_3\times \Z_3$-symmetric potential is automatically symmetric
under $(\Z_3 \times \Z_3) \rtimes \Z_2$, see explicit expressions below.
However, since it appears as an abelian subgroup
of a finite realizable group, it must be included into consideration.
Trying to impose any other abelian Higgs-family symmetry group on the 3HDM potential
unavoidably makes it symmetric under a continuous group.

Let us first see what order the finite (non-abelian) group $G$ can have.
We note that the orders of all abelian groups in (\ref{abelian}) have only two prime divisors: 2 and 3.
Thus, by Cauchy's theorem, the order of the group $G$ can also have only these two prime
divisors: $|G| = 2^a 3^b$. Then according to the Burnside's $p^aq^b$-theorem the group $G$ is solvable
(Theorem 7.8 in \cite{Isaacs}), and this means that $G$ contains a normal abelian subgroup,
which belongs, of course, to the list (\ref{abelian}).

In order to proceed further, we need to prove that one can
in fact find a normal maximal (that is, self-centralizing) abelian subgroup of $G$,
a property which is not generic to solvable groups but which holds in our case.

\subsection{Existence of a normal abelian self-centralizing subgroup}\label{subsection-self-centralizing}

Suppose $A  <  G$ is a normal abelian subgroup,
whose existence follows from the solvability of $G$.
In this subsection we prove that even if it is not self-centralizing, i.e. $A  <  C_G(A)$, then
there exists another abelian subgroup $B  >  A$, which is normal and self-centralizing in $G$.

\begin{figure}[!htb]
   \centering
\includegraphics[height=5cm]{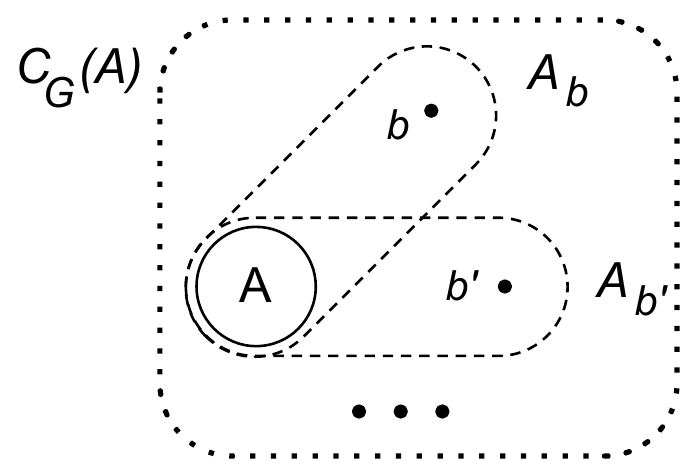}
\caption{Illustration of $C_G(A)$ and some of its subgroups.}
   \label{fig-CGA}
\end{figure}
Suppose that $A  <  C_G(A)$.
Then for every $b \in C_G(A)\setminus A$, the group $A_b = \langle A, b\rangle$ is an abelian subgroup of $G$,
which properly contains $A$. 
Fig.~\ref{fig-CGA} should help visualize embedding of various abelian subgroups of this kind in $C_G(A)$.
Note that $C_G(A)$ can be non-abelian.
There are two possibilities compatible with the list (\ref{abelian}):
\begin{itemize}
\item[(i)]
$A = \Z_2$, and then $A_b$ can be either $\Z_2\times \Z_2$ or $\Z_4$,
\item[(ii)]
$A = \Z_3$, and then $A_b = \Z_3\times\Z_3$.
\end{itemize}
Thus $C_G(A)$ is either a $2$-group or a $3$-group. Below we assume that $p=2$ if $C_G(A)$ is a $2$-group, and $p=3$
if $C_G(A)$ is a $3$-group.

Since $C_G(A)$ is a $p$-group, it is nilpotent, and according to discussion in section~\ref{subsection-nilpotent},
it possesses a normal maximal abelian subgroup $B$ 
(which of course can be represented as $A_b$ for some $b$),
while $B$ properly includes $A=\Z_p$: $A < B \leq C_G(A)$. 
In particular, $B$ is self-centralizing in $C_G(A)$, so according to our discussion in section~\ref{subsection-maximal-normal-abelian},
the factor group $C_G(A)/B$ is a subgroup of $Aut(B)$.
If $B=C_G(A)$, then $C_G(A)$ is abelian and, being a centralizer of a normal subgroup, it is normal in $G$.
Clearly $B\leq C_G(B)\leq C_G(A)=B$, therefore $C_G(A)$ is the desired normal abelian self-centralizing subgroup of $G$.

Assume now that $B\not=C_G(A)$:
\be
A < B = \underbrace{C_{C_G(A)}(B)}_{=C_G(B)} < C_G(A) < G\,.
\ee
The illustration in Fig.~\ref{fig-CGA} refers to this case.
Since $B$ is an abelian subgroup of $G$,
it must be in list~\eqref{abelian}. So, either $B=\Z_p\times \Z_p$ or
$B=\Z_{p^2}$ (the last case occurs only if $p=2$), and in any of these cases we obtain $\vert B\vert=p^2$.
Now, recall that $C_G(A)$ is a $p$-group, and so is $C_G(A)/B$.
If $B=\Z_p\times \Z_p$, then $C_G(A)/B$ is a $p$-subgroup of
$GL_2(p)$, in particular, $\vert C_G(A)/B\vert=p$. If $B=\Z_{p^2}$, then $C_G(A)/B$ is a $p$-subgroup of
$Aut(\Z_{p^2})$. Since $\varphi(p^2)=p(p-1)$, it follows that $\vert C_G(A)/B\vert=p$. So in any case we have $\vert
C_G(A)\vert=p^3$.

Now the arguments depend on $p$.
\begin{itemize}
\item
In the case $p=2$,
we have that $C_G(A)$ is a nonabelian group of order $8$. Thus $C_G(A)$ is either dihedral group $D_8$
or the quaternion group $Q_8$. If $C_G(A)$
is dihedral, then it possesses the unique (and hence characteristic) subgroup $H=\Z_4$, so $H$ is the desired normal
self-centralizing subgroup of $G$. If $G=Q_8$ is quaternion then, as we describe in Section~\ref{attemptingQ8}, trying to
impose a $Q_8$ symmetry group on the 3HDM potential
will result in a potential symmetric under a continuous group. Thus, this situation cannot happen if we search for
finite realizable groups $G$.
Note that this feature is purely calculational and
does not rely on the existence of a  normal maximal abelian subgroup which we prove here.
\item
In the case $p=3$,
we have that $C_G(A)$ is a nonabelian group of order $p^3=27$ and exponent $3$, i.e. for every $g\in C_G(A)$ we have $g^3=1$.
It is nonabelian and cannot contain elements of order $9$ because (\ref{abelian}) does not contain abelian groups of orders $9$ or $27$. 

In this case we do not yet know whether $B$ is normal in $G$, but it is definitely normal in its own normalizer
$B \lhd N_G(B) \leq G$. Moreover $C_G(A)\leq N_G(B)$, since $B$ is normal in $C_G(A)$. 
These relations are visualized by the following relations:
\be
B \lhd C_G(A) \le N_G(B) \le G < PSU(3)\,.
\ee
We can then consider the
factor group  $N_G(B)/B$. We know that $B=\Z_3\times \Z_3$ is a maximal abelian group in $PSU(3)$, \cite{abelianNHDM};
therefore it is self-centralizing in $PSU(3)$ and, consequently, in $G$
and in its subgroup $N_G(B)$.
Then, in particular, we have that
$N_G(B)/B$ is a subgroup of $Aut(B)=GL_2(3)$. Moreover, the analysis which will be exposed in detail in
Section~\ref{sectionZ3Z3} allows us to state that $N_{PSU(3)}(B)/B=SL_2(3)$, so $N_G(B)/B$ is a subgroup of
$SL_2(3)$.
We show in Section~\ref{sectionZ3Z3} that one cannot use elements of order 3 from $SL_2(3)$ because the potential will
then become invariant under a continuous symmetry group. Therefore, $N_G(B)/B$ cannot have elements of
order $3$, which implies that $B$ is a Sylow $3$-subgroup of $N_G(B)$.
The same statement holds for every group that lies ``between'' $N_G(B)$ and $B$, in particular, to $C_G(A)$. This contradicts the fact that
$\vert C_G(A):B\vert=3$ and $C_G(A)\leq N_G(B)$. So this case is impossible.
\end{itemize}
Summarizing the group-theoretic part of our derivation, we proved that any finite group $G$
which can be realized as a Higgs-family symmetry group in 3HDM is solvable,
and in addition it contains a normal self-centralizing abelian subgroup $A$
belonging to the list (\ref{abelian}).
Then, according to (\ref{structure-G}) the group $G$ can be constructed
as an extension of $A$ by a subgroup of $Aut(A)$.

This marks the end of the group-theoretic part of our analysis.
We now need to check all the five candidates for $A$,
whose explicit realization were already given in \cite{abelianNHDM},
and by means of direct calculations see which extension can work in 3HDM.

\section{Detecting continuous symmetries}\label{section-continuous}

Before we embark on analyzing each particular abelian group and its extensions,
let us discuss an important issue.
In this paper, we focus on discrete symmetries of the scalar sector in 3HDM.
The symmetry groups we study must be realizable, that is, we need to prove that
a potential symmetric under a finite group $G$ is not symmetric under any larger group
containing $G$. In particular, we must prove
that a given $G$-symmetric potential does not have any continuous symmetry.

In principle, it would be desirable to derive a basis-invariant criterion
for existence or absence of a continuous symmetry. Such condition is known for 2HDM, \cite{2HDMsymmetries,ivanov2HDM},
while for the more than two doublets a necessary and sufficient condition is still missing.
However, in certain special but important cases it is possible to derive a {\em sufficient} condition for absence of any continuous symmetry.
Since this method relies on the properties of the orbit space in 3HDM, we start by briefly describing it.

\subsection{Orbit space in 3HDM}

The formalism of representing the space of electroweak-gauge orbits of Higgs fields
via bilinears was first developed for 2HDM, \cite{2HDMbilinears,2HDMsymmetries,ivanov2HDM},
and then generalized to $N$ doublets in \cite{ivanovNHDM}. Below we focus on the 3HDM case.

The Higgs potential depends on the Higgs doublets via their gauge-invariant bilinear combinations
$\phi^\dagger_a \phi_b$, $a,b = 1,2,3$.
These bilinears can be organized into the following real scalar $r_0$ and real vector $r_i$, $i = 1,\dots,8$:
\bea
&& r_0 = {(\phi_1^\dagger\phi_1) + (\phi_2^\dagger\phi_2) + (\phi_3^\dagger\phi_3)\over\sqrt{3}}\,,\quad
r_3 = {(\phi_1^\dagger\phi_1) - (\phi_2^\dagger\phi_2) \over 2}\,,\quad
r_8 = {(\phi_1^\dagger\phi_1) + (\phi_2^\dagger\phi_2) - 2(\phi_3^\dagger\phi_3) \over 2\sqrt{3}}\,,\quad
\nonumber\\
&&r_1 = \Re(\phi_1^\dagger\phi_2)\,,\quad r_2 = \Im(\phi_1^\dagger\phi_2)\,,\quad
r_4 = \Re(\phi_3^\dagger\phi_1)\,,\nonumber\\[2mm] &&r_5 = \Im(\phi_3^\dagger\phi_1)\,,\quad
r_6 = \Re(\phi_2^\dagger\phi_3)\,,\quad r_7 = \Im(\phi_2^\dagger\phi_3)\,.\label{ri}
\eea
The last six components can be grouped into three ``complex coordinates'':
\be
r_{12} = (\phi_1^\dagger\phi_2) = r_1 + i r_2\,,
\quad r_{45} = (\phi_3^\dagger\phi_1) = r_4 + i r_5\,,
\quad r_{67} = (\phi_2^\dagger\phi_3) = r_6 + i r_7\,.\label{complexcoordinates}
\ee
It is also convenient to define the normalized coordinates $n_i = r_i/r_0$.
The orbit space of the 3HDM is then represented by an algebraic manifold lying in the $1+8$-dimensional euclidean space
of $r_0$ and $r_i$ and is defined by the following (in)equalities, \cite{ivanovNHDM}:
\be
r_0 \ge 0\,,\quad \vec n^2 \le 1\,, \quad \sqrt{3}d_{ijk} n_i n_j n_k = {3 \vec n^2 - 1\over 2}\,,\label{orbitspace}
\ee
where $d_{ijk}$ is the fully symmetric $SU(3)$ tensor.
It can also be derived that $|\vec n|$ is bounded from below:
\be
\vec n^2 = \alpha\,,\quad {1\over 4} \le \alpha \le 1\,.\label{alpha}
\ee
The value of $\alpha$ parametrizes $SU(3)$-orbits inside the orbit space.
In particular, we will use this relation below when substituting $r_3^2+r_8^2$ by
$\alpha r_0^2 - |r_{12}|^2 - |r_{45}|^2 - |r_{67}|^2$.

Any $U(3)$ transformation in the space of doublets $\phi_1$, $\phi_2$, $\phi_3$
leaves $r_0$ invariant and
induces an $SO(8)$ rotation of the vector $r_i$.
Note that this map is not surjective, namely not every $SO(8)$ rotation of $r_i$ can be induced
by a $U(3)$ transformation in the space of doublets.
Therefore, unlike in 2HDM, we do not expect the orbit space of 3HDM to be $SO(8)$-symmetric,
and the last condition in (\ref{orbitspace}) stresses that.

\begin{figure}[!htb]
   \centering
\includegraphics[height=6cm]{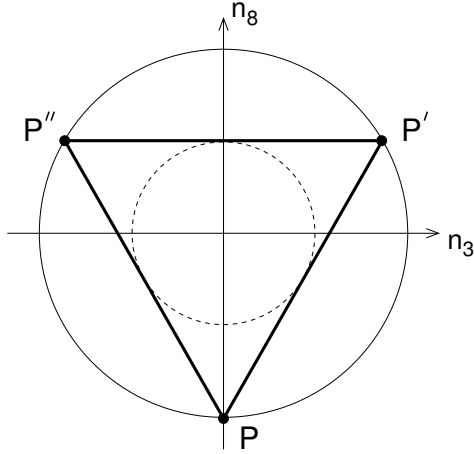}
\caption{The orbit space of 3HDM in the $(n_3,n_8)$-subspace (all other $n_i=0$). The outer and inner
circles correspond to $|\vec n|=1$ and $|\vec n|=1/2$, respectively.}
   \label{fig-n38}
\end{figure}
Let us take a closer look at the $(n_3,n_8)$-subspace. It follows from (\ref{orbitspace})
that the orbit space intersects this plane along the equilateral triangle shown in Fig.~\ref{fig-n38}.
Its vertices $P$, $P'$, $P''$ lie on the ``neutral'' manifold, which satisfy the condition $\vec n^2 = 1$
and which would correspond to the neutral vacuum
if the minimum of the potential were located there, while the line segments joining them correspond to the
charge-breaking vacuum, see details in \cite{ivanovNHDM}.
The orbit space in this plane clearly lacks the rotational symmetry and has only the symmetries of the equilateral triangle.

\subsection{Absence of continuous symmetries}

The convenience of the formalism of bilinears is that the most general Higgs potential becomes
a quadratic form in this space:
\be
\label{potential:again}
V = - M_0 r_0 - M_i r_i + {1 \over 2}\Lambda_{00} r_0^2 + \Lambda_{0i} r_0 r_i + {1 \over 2}\Lambda_{ij} r_i r_j\,.
\ee
The real symmetric matrix $\Lambda_{ij}$ has eight real eigenvalues (counted with multiplicity).
In order for the potential to be symmetric under a continuous group of transformations,
$\Lambda_{ij}$ must have eigenvalues of multiplicities $>1$. Note that any statement about eigenvalues of
$\Lambda_{ij}$ is basis-invariant and therefore it can be checked in any basis.
Furthermore, if we find a basis in which $\Lambda_{ij}$ has a block-diagonal form,
and if eigenvalues from different blocks are distinct, then a continuous symmetry requires that
each block is either invariant under this symmetry, or contains eigenvalues with multiplicity $> 1$.

Let us consider an important special case of this situation.
Suppose that the potential has no terms of type $(\phi_a^\dagger \phi_a)(\phi_b^\dagger \phi_c)$, where
$a, b, c$ are all distinct. This implies the absence of terms $r_{0,3,8}r_{1,2,4,5,6,7}$,
and the block-diagonal form of $\Lambda_{ij}$, in which two blocks correspond to the $(r_3,r_8)$ subspace
and to its orthogonal complement.
Suppose also that the eigenvalues of $\Lambda_{ij}$ in the $(r_3,r_8)$ subspace are distinct from those
in the orthogonal complement. It follows then that any possible continuous symmetry must act {\em trivially}
in the $(r_3,r_8)$ subspace, because the orbit space here lacks the rotational invariance.
However, if $r_0$, $r_3$, and $r_8$ are fixed, then $\phi_1^\dagger \phi_1$, $\phi_2^\dagger \phi_2$, 
and $\phi_3^\dagger \phi_3$ are also fixed. So, the doublets do not mix, 
and the possible continuous symmetry group can only be a subgroup
of the group of pure phase rotations, which were studied in \cite{abelianNHDM}.

If in addition it is known that a given potential is not symmetric under continuous phase rotations,
then we conclude that it does not have any continuous symmetry from $PSU(3)$.
It turns out that all the cases of various finite symmetry groups we consider below, except the last one, are of this type.
Since the arguments of this section provide a sufficient condition for absence of continuous symmetries,
they guarantee
that the corresponding potentials can have only finite symmetry groups.
Absence of a continuous symmetry in the very last case will be proved separately.

\section{Possible extensions: the torus chain}\label{section-torus}

We now check all the candidates for $A$ from the list (\ref{abelian})
and see which extension can work in 3HDM.
In this section we will deal with the first four groups from the list, which arise as
subgroups of the maximal torus; the last group will be considered later.
For each group $A$, we use its explicit realization given in \cite{abelianNHDM} as a group of rephasing transformations,
and then we search for additional transformations from $PSU(3)$
with the desired multiplication properties.

\subsection{Representing elements of $PSU(3)$}

Before we start analysis of each case, let us make a general remark on how we describe the elements of $PSU(3)$.
Using the bar notation for the canonical homomorphism $SU(3) \to PSU(3)$, we
denote $\bar H  <  PSU(3)$ if its full preimage in $SU(3)$ is $H$. Denoting the center of $SU(3)$ as
$Z = Z(SU(3)) \simeq \Z_3$, we have $Z = \{1,\, z,\, z^2\}$, where
\be
z = \mathrm{diag}(\omega,\omega,\omega)\,,\quad \omega = e^{2\pi i/3}\,.
\ee
The elements of the group $H$ ($a, b, \ldots \in H$) will be written as $3\times 3$ matrices from $SU(3)$.
The elements of $\bar H$ ($\bar a, \bar b, \dots \in \bar H$) are the corresponding cosets of $Z$ in $H$. Explicit manipulation with these cosets
is inconvenient, therefore in our calculation we represent an element $\bar a \in PSU(3)$
by any of the three representing elements from $SU(3)$: $a$, $az$, or $az^2$.
We will usually choose $a$ and then prove that this representation is faithful
(does not depend on the choice of representing element).

\subsection{Extending $\Z_2$ and $\Z_3$}\label{subsection-D6}

The smallest group from the list is $A = \Z_2$, whose automorphism group is $Aut(\Z_2) = \{1\}$, so that $G = \Z_2$.
This case was already considered in \cite{abelianNHDM}.

The next possibility is $A = \Z_3$, whose $Aut(\Z_3) = \Z_2$. The only non-trivial case to be considered is $G/A = \Z_2$,
which implies that $G$ can be either $\Z_6$ or $D_6 \simeq S_3$, the symmetry group of the equilateral triangle.
The former can be disregarded because it does not appear in the list (\ref{abelian}), thus we focus only on the $D_6$ case.

\subsubsection{Constructing $D_6$}
The group $D_6$ is generated by two elements $a, b$ with the following relations: $a^3 = 1$, $b^2 = 1$, $ab = ba^2$.
Following \cite{abelianNHDM}, we represent the $\Z_3$ group by phase rotations:
\be
a = \mathrm{diag}(\omega,\omega^2,1)\,.
\ee
There are in fact three such groups which differ only by the choice of the doublet which is fixed.
However their generators, $a$, $az$, and $az^2$, differ only by a transformation from the center,
and therefore all of them correspond to the same generator $\bar a$ from $PSU(3)$.
It is straightforward to check that selecting $a$ to represent $\bar a$ is a faithful representation.

The explicit solution of the matrix equation $ab = ba^2$ shows that $b \in SU(3)$ must be of the form
\be
b = \left(\begin{array}{ccc} 0 & e^{i\delta} & 0 \\ e^{-i\delta} & 0 & 0 \\ 0 & 0 & -1 \end{array}\right)\,,\label{bZ3}
\ee
with an arbitrary $\delta$.
The choice of the mixing pair of doublets ($\phi_1$ and $\phi_2$ in this case) is fixed by the choice of invariant doublet in $a$.

The fact that $b$ is not uniquely defined means that there exists not a single $D_6$ group
but a whole family of $D_6$ groups parametrized by the value of $\delta$.
Below, when checking whether a potential is $D_6$ symmetric, 
we will need to check its invariance under all possible $D_6$'s from this family.

The generic $\Z_3$-symmetric potential contains the part invariant under any phase rotation
\be
V_0 =  - \sum_{1\le i \le 3} m_i^2(\phi_i^\dagger \phi_i) + \sum_{1 \le i\le j \le 3} \lambda_{ij} (\phi_i^\dagger \phi_i)(\phi_j^\dagger \phi_j)\nonumber\\
+ \sum_{1 \le i < j \le 3} \lambda'_{ij} (\phi_i^\dagger \phi_j)(\phi_j^\dagger \phi_i)\,,\label{Tsymmetric}
\ee
and the following additional terms
\be
V_{\Z_3} = \lambda_1(\phi_2^\dagger \phi_1)(\phi_3^\dagger \phi_1) + \lambda_2(\phi_1^\dagger \phi_2)(\phi_3^\dagger \phi_2)
+ \lambda_3 (\phi_1^\dagger \phi_3)(\phi_2^\dagger \phi_3) + h.c.\label{VZ3}
\ee
with complex $\lambda_1,\, \lambda_2,\,\lambda_3$.
At least two of them must be non-zero, otherwise the potential will be symmetric
under a continuous group of Higgs-family transformations, \cite{abelianNHDM}.
Let us denote their phases by $\psi_1$, $\psi_2$, and $\psi_3$, respectively.
If the parameters of $V_0$ satisfy
\be
m_{11}^2 = m_{22}^2\,, \ \lambda_{11} = \lambda_{22}\,,\
\lambda_{13} = \lambda_{23}\,,\ \lambda'_{13} = \lambda'_{23}\,,\label{conditionV0}
\ee
and if, in addition, $|\lambda_1| = |\lambda_2|$,
then the whole potential becomes symmetric under one particular $D_6$ group
constructed with $b$ in (\ref{bZ3}) with the value of $\delta = (\psi_2 - \psi_1 + \pi)/3 + 2\pi k/3$.
The extra freedom given by $2\pi k/3$ corresponds to three order-two elements of $D_6$:
$b$, $ab$, $a^2b$. We opt to define $b$ by setting $k=0$.
Alternatively, we can be compactly write the condition as
\be
3\delta = \pi - \psi_1 + \psi_2\,. \label{conditiondeltaZ3}
\ee
To summarize, the criterion of the $D_6$ symmetry of the potential is that, after a possible doublet relabeling, 
conditions (\ref{conditionV0}) and (\ref{conditiondeltaZ3}) are satisfied.

Let us also note that when constructing the group $D_6$ we could have searched for $b$ satisfying
not $ab = ba^2$ but $ab = ba^2 \cdot z^p$, with $p = 1,2$. Solutions of this equation exist, but they do not
lead to any new possibilities.
Indeed, let us introduce $a' = az^p$. Then, we get $a'b = ba^{\prime 2}$. Thus, we get the same equation for $b$
as before, up to a cyclic permutation of doublets, the possibility which we already took into account.

\subsubsection{Proving that $D_6$ is realizable}\label{section-proof-of-realizability}

This construction allows us to write down an example of the $D_6$-symmetric potential:
it is $V_0$ restricted by conditions (\ref{conditionV0}) 
plus $V_{\Z_3}$ in (\ref{VZ3}) subject to $|\lambda_1| = |\lambda_2|$.
In order to show that $D_6$ is realizable, we need to demonstrate that 
this potential is not symmetric under any larger Higgs-family transformation group.

This proof is short and contains two steps.
First, we note that the conditions described in section~\ref{section-continuous} are fulfilled:
the $(r_3,r_8)$-subspace does not couple to its orthogonal complement via $\Lambda_{ij}$,
and that the eigenvalues in these two subspaces are defined by different sets of free parameters.
The extra terms (\ref{VZ3}) guarantee that there is only finite group of phase rotations, the group $\Z_3$.
Therefore, 
the sufficient conditions described in section~\ref{section-continuous} are satisfied, and
the generic $D_6$-symmetric potential has no continuous symmetry.

Second, we need to show that the generic $D_6$-symmetric potential has no higher discrete symmetries.
This is proved by the simple observation that all other finite groups to be discussed below
which could possibly contain $D_6$ lead to {\em stronger} restrictions on the potential than (\ref{conditionV0})
and $|\lambda_1| = |\lambda_2|$. Therefore, not satisfying those stronger restrictions
will yield a potential symmetric only under $D_6$.

\subsubsection{Including antiunitary transformations}

Any generalized-$CP$ (antiunitary) transformation acting on three doublets
is of the form
\be
J' = c \cdot J\,,\quad c \in PSU(3)\,.
\ee
Here $J$ is the operation of hermitean conjugation of the doublets.
If $G$ is the symmetry group of unitary transformations, then it is normal in $\langle G, J'\rangle$,
and $J'$ induces automorphisms in $G$.
So, when we search for $J'$, we require that
\be
(J')^2 \in G\,,\quad (J')^{-1} a J' \in G\,,\label{closure}
\ee
where $a$ generically denotes the generators of $G$.
If such a transformation is found, the group is extended from $G$ to $G\rtimes \Z_2^*$,
where asterisk on the group indicates that its generator is antiunitary.

Note the crucial point of our method: when extending $G$ by an antiunitary transformation, we require that the
unitary transformation symmetry group remains $G$.
The logic is simple. If we start with a realizable group $G$ of unitary transformations but
do not impose condition (\ref{closure}), we will end up with a potential being symmetric under
$\tilde G\rtimes \Z_2^*$, with $\tilde G > G$.
But at the end of this paper we will have a complete list of all finite realizable symmetry groups
of unitary transformations, and this list will contain $\tilde G$ anyway. So, this possibility
is not overlooked but will be studied in its due time after construction of $\tilde G$.

Now, turning to extension of $D_6$ by an antiunitary symmetry,
we first note that the resulting group $D_6 \rtimes \Z_2^*$ is a non-abelian group of order 12
containing a normal subgroup $D_6$. Among the three non-abelian groups of order 12,
there exists only one group, namely $D_6 \times \Z_2^*$, with a subgroup $D_6$ (which is automatically normal
because all subgroups of index 2 are normal). This fact can also be proved
in a more general way without knowing the list of groups of order 12.
Note that it contains, among other, the subgroup $\Z_6^*$;
its presence does not contradict the list (\ref{abelian}) because that list refers
only to the groups of unitary transformations.

Next, let us denote the generator of $\Z_2^*$ by $J' = c J$. Since $J'$ centralizes the entire $D_6$,
it follows that $(J')^{-1} a J' = a$, $(J')^{-1} b J' = b$, and $(J')^2 = cJcJ = cc^* = 1$.
The matrix $c$ satisfying these conditions must be of the form 
\be
c =  \left(\begin{array}{ccc}
0 & e^{i\gamma} & 0 \\
e^{i\gamma} & 0 & 0 \\
0 & 0 & -e^{-2i\gamma}
\end{array}\right)\,,
\label{cgamma}
\ee
with arbitrary $\gamma$.
Requiring the potential to stay invariant under $J'$, we obtain the following conditions on $\gamma$:
$6\gamma = -2(\psi_1 + \psi_2) = 2\psi_3$. 
Therefore, if the following {\em extra} condition is fulfilled, 
\be
2(\psi_1 + \psi_2 +\psi_3) = 0\,.\label{conditionD6Z2}
\ee
the $D_6$-invariant potential becomes symmetric under the group $D_6 \times \Z_2^*$.
If this condition is not satisfied, the symmetry group remains $D_6$ even in the case when
antiunitary transformations are allowed.
We conclude that both $D_6$ and $D_6 \times \Z_2^*$ are realizable in 3HDM.

It is interesting to note that if we set $\lambda_3 = 0$, then the potential would still be invariant under $D_6$.
However in this case it becomes symmetric under $J'$ with $6\gamma = -2(\psi_1 + \psi_2)$,
without any extra condition on $\psi_1$ and $\psi_2$,
and the potential becomes automatically invariant under $D_6 \times \Z_2^*$.
So, we conclude that the fact that $D_6$ is still realizable even if anti-unitary transformations are included is due
to the special feature of the $\Z_3$-symmetry: we have three, not two terms in the $\Z_3$-symmetric potential,
and it is the third term that prevents an automatic anti-unitary symmetry.

\subsection{Extending $\Z_4$}

Let us now take $A = \Z_4$ generated by $a$. Then $Aut(\Z_4) = \Z_2$, so that $G = \Z_4\arbitraryext \Z_2$ generated by $a$
and some $b \not \in \Z_4$.
The two non-abelian possibilities for $G$ are
the dihedral group $D_8$ representing symmetries of the square, and the quaternion group $Q_8$.
In both cases $b^{-1}ab = a^3$, with the only difference that $b^2 = 1$ for $D_8$
while $b^2 = a^2$ for $Q_8$. Note that extension leading to the dihedral group is split, $D_8 = \Z_4 \rtimes \Z_2$, while
$Q_8$ is not.

\subsubsection{Constructing $D_8$}\label{constructingD8}

Representing $a$ by phase rotations $a = \mathrm{diag}(i,-i,1)$, we find that $b$ satisfying these conditions is again
of the form (\ref{bZ3}) with arbitrary $\delta$.
However now we do not have the freedom to choose the pair of doublets which are mixed by $b$: this pair is fixed by $a$.
Also, unlike the $\Z_3$ case, the matrix equation $ab = ba^3\cdot z$ does not have solutions for $b \in SU(3)$.

The $\Z_4$-symmetric potential (for this choice of $a$) is $V_0 + V_{\Z_4}$, where
\be
V_{\Z_4} = \lambda_1 (\phi_3^\dagger \phi_1)(\phi_3^\dagger \phi_2) + \lambda_2 (\phi_1^\dagger \phi_2)^2 + h.c. \label{VZ4}
\ee
The phases of $\lambda_1$ and $\lambda_2$ are, as usual, denoted as $\psi_1$ and $\psi_2$, respectively.
Upon $b$, the first term here remains invariant, while the second term transforms as
\be
(\phi_1^\dagger \phi_2)^2 \mapsto e^{-4i\delta}(\phi_2^\dagger \phi_1)^2\,.
\ee
This means that the potential (\ref{VZ4}) is {\em always} symmetric under (\ref{bZ3}) provided that we choose
\be
\delta = \psi_2/2\,,\label{conditiondeltaZ4}
\ee
Therefore, in order to get a $D_8$-symmetric potential we only require that $V_0$ satisfies conditions (\ref{conditionV0}).
The proof that $D_8$ is realizable (as long as only unitary transformations are concerned) follows along the same lines as
in section \ref{section-proof-of-realizability}.

\subsubsection{Including antiunitary transformations}\label{subsubsection-D8Z2}

In \cite{abelianNHDM} we found that exactly the same conditions, namely (\ref{conditionV0}) and (\ref{conditiondeltaZ4}),
must be satisfied for existence of an antiunitary transformation commuting with
the elements of $\Z_4$. 
This transformation is again $J' = cJ$, where $c$ is given by (\ref{cgamma}) with $6\gamma = 2 \psi_1$,
and it commutes with all elements of $D_8$.
Therefore, if we include antiunitary transformations,
we automatically get the group $D_8 \times \Z_2^*$, while $D_8$ becomes non-realizable.
Note that the resulting group does not contain $\Z_8^*$. Indeed, we showed in \cite{abelianNHDM}
that imposing $\Z_8^*$ symmetry group leads to a potential with continuous symmetry.

\subsubsection{Attempting at $Q_8$}\label{attemptingQ8}

Solving matrix equations $ab = ba^3$ and $b^2 = a^2$, we get the following form of $b$:
\be
b(Q_8) = \left(\begin{array}{ccc} 0 & e^{i\delta} & 0 \\  - e^{-i\delta} & 0 & 0 \\ 0 & 0 & 1 \end{array}\right)\,.\label{bQ8}
\ee
By checking how $V_{\Z_4}$ in (\ref{VZ4}) transforms under it, we find that the first term simply changes
its sign. The only way to make the potential symmetric under $Q_8$ is to set $\lambda_1 = 0$.
But then we know from \cite{abelianNHDM} that the potential
becomes invariant under a continuous group of phase rotations.
Therefore, $Q_8$ is not realizable.

\subsection{Extending $\Z_2 \times \Z_2$}

If $A = \Z_2 \times \Z_2$, then $Aut(\Z_2\times \Z_2) = GL_2(2) = S_3$. The group
$\Z_2 \times \Z_2$ can be realized as the group of independent sign flips of the three doublets with
generators $a_1 = \mathrm{diag}(1,-1,-1)$ (equivalent to the sign flip of the first doublet)
and $a_2 = \mathrm{diag}(-1,1,-1)$ (equivalent to the sign flip of the second doublet),
so that $a_1a_2$ is equivalent to the sign flip of the third doublet.
The potential symmetric under this group contains $V_0$ and additional terms
\be
V_{\Z_2 \times \Z_2} = \tilde \lambda_{12} (\phi_1^\dagger \phi_2)^2 +
\tilde \lambda_{23} (\phi_2^\dagger \phi_3)^2 +
\tilde \lambda_{31} (\phi_3^\dagger \phi_1)^2 + h.c.\label{VZ2Z2}
\ee
with at least two among coefficients $\tilde \lambda_{ij}$ being non-zero.
The coefficients can be complex; as usual we denote their phases as $\psi_{ij}$.
This model is also known as the Weinberg's 3HDM, \cite{weinberg3HDM}.

The non-abelian finite group $G$ can be constructed as extension of $A$ by $\Z_2$, by $\Z_3$, or by~$S_3$.

\subsubsection{Extension $(\Z_2 \times \Z_2)\arbitraryext \Z_2$}

Consider first the extension $(\Z_2 \times \Z_2)\arbitraryext \Z_2$.
The only extension leading to a non-abelian group is $(\Z_2 \times \Z_2)\arbitraryext \Z_2=D_8$, 
and we already proved that this group is realizable.
Nevertheless, we prefer to explicitly work it out to see the reduction of free parameters.

The element $b$ which we search for must act on $\{a_1,a_2,a_1a_2\}$ as a transposition of any pair.
In addition, $b^2 \in \Z_2 \times \Z_2$.
It does not matter which pair of generators is transposed, as this choice can be changes by renumbering the doublets.
So, we take $b$ such that $b^{-1}a_1 b = a_2$ and $b^{-1}a_2 b = a_1$. Then, $b^2$ can be either 1 or $a_1a_2$,
because choices $b^2 = a_1$ or $a_2$ lead to inconsistent relations. Indeed, if we assume $b^2=a_1$,
then
$$
a_2 = b^{-1}a_1 b = b^{-1}b^2 b = b^2 = a_1\,,
$$
which is a contradiction.
In both cases ($b^2 = 1$ and $b^2 = a_1a_2$) we get the group $D_8$.
Even more, we get the {\em same} $D_8$ group:
if $b^2 = a_1a_2$, then $b' = b a_1$ satisfies $b^{\prime 2} = 1$,
while its action on $a_1$ and $a_2$ remains the same.
So, it is sufficient to focus on the $b^2=1$ case only.

Again, explicitly solving the matrix equations, we get $b$ of the form (\ref{bZ3}) with arbitrary $\delta$.
Then, we check how the potential (\ref{VZ2Z2}) changes upon $b$ and find that we need to set
\be
4\delta = 2\psi_{12}\,,\quad 2\delta = -(\psi_{23} + \psi_{31})\,,\quad |\tilde\lambda_{23}| = |\tilde\lambda_{31}|\,.
\ee
Equations on the phase $\delta$ can be satisfied if
\be
2(\psi_{12} + \psi_{23} + \psi_{31}) = 0\quad \Leftrightarrow \quad
\Im(\tilde\lambda_{12}\tilde\lambda_{23}\tilde\lambda_{31}) = 0\,.\label{conditionD8}
\ee
So, if: (1) this condition is satisfied, (2) two among $|\tilde\lambda_{ij}|$ are equal, (3) condition on $V_0$ (\ref{conditionV0})
is satisfied, then the potential is $D_8$-symmetric.
Note also that if $\tilde \lambda_{12} = 0$ (which we are allowed to consider because (\ref{VZ2Z2}) contains three rather than two terms),
then condition on the phases is not needed.

It might seem that these conditions on the potential to make it $D_8$-symmetric are more restrictive than
in the $\Z_4$ extension we studied above. However note that the $\Z_2\times \Z_2$-symmetric potential (\ref{VZ2Z2})
has six free parameters, and we placed two conditions to reduce the number of free parameters in the $D_8$ potential to
four (apart from $V_0$). On the other hand, (\ref{VZ4}) had only four from the beginning, and without any restriction
this number survives. Therefore we have the same number of degrees of freedom when constructing $D_8$
in either way.

\subsubsection{Constructing $(\Z_2 \times \Z_2)\rtimes Z_3 = T$}

The extension by $\Z_3$ is necessarily split, $(\Z_2 \times \Z_2)\rtimes \Z_3$, leading to the group $T \simeq A_4$,
the symmetry group of the tetrahedron.
To construct it, we need $b$ such that $b^3 = 1$ with the property that $b$ acts on $\{a_1,a_2,a_1a_2\}$ by cyclic permutations.
Fixing the order of permutations by $b^{-1}a_1 b = a_2$, we find that $b$ must be of the form
\be
b = \left(\begin{array}{ccc} 0 & e^{i\delta_1} & 0 \\ 0 & 0 & e^{i\delta_2} \\ e^{-i(\delta_1+\delta_2)} & 0 & 0 \end{array}\right)\,,
\ee
with arbitrary $\delta_1$, $\delta_2$.
It then follows that if coefficients in (\ref{VZ2Z2}) satisfy
\be
|\tilde \lambda_{12}| = |\tilde \lambda_{23}| = |\tilde \lambda_{31}|\,,\label{conditionT}
\ee
then $V_{\Z_2 \times \Z_2}$ is symmetric under one particular $b$ with
\be
\delta_1 = {2\psi_{12} - \psi_{31} - \psi_{23} \over 6}\,,\quad
\delta_2 = {2\psi_{23} - \psi_{31} - \psi_{12} \over 6}\,.\nonumber
\ee
Then, by a rephasing transformation one also make the phases of all $\tilde \lambda_{ij}$ equal and bring (\ref{VZ2Z2}) to the following form
\be
V_T = \tilde\lambda  \left[(\phi_1^\dagger\phi_2)^2 + (\phi_2^\dagger\phi_3)^2 + (\phi_3^\dagger\phi_1)^2\right] + h.c.\label{VT}
\ee
with a complex $\tilde \lambda$. In this form, the parameters $\delta_1=\delta_2 = 0$, 
and the matrix $b$ is just the cyclic permutation of the doublets.
In addition, the symmetry under $b$ places stronger conditions on the parameters of $V_0$, 
so that the most general $V_0$ satisfying them is
\bea
V_0 &=& -m^2 \left[(\phi_1^\dagger\phi_1) + (\phi_2^\dagger\phi_2) + (\phi_3^\dagger\phi_3)\right] +
\lambda \left[(\phi_1^\dagger\phi_1) + (\phi_2^\dagger\phi_2) + (\phi_3^\dagger\phi_3)\right]^2  \label{V0Z2Z2Z3}\\
&& + \lambda' \left[(\phi_1^\dagger\phi_1)(\phi_2^\dagger\phi_2) + (\phi_2^\dagger\phi_2)(\phi_3^\dagger\phi_3) + 
(\phi_3^\dagger\phi_3)(\phi_1^\dagger\phi_1)\right]
+ \lambda'' \left(|\phi_1^\dagger\phi_2|^2 + |\phi_2^\dagger\phi_3|^2 + |\phi_3^\dagger\phi_1|^2\right)\,.\nonumber
\eea

\subsubsection{Constructing $(\Z_2 \times \Z_2)\rtimes S_3 = O$}

The last extension, $(\Z_2 \times \Z_2)\arbitraryext S_3$,
is also split, otherwise we would obtain $\Z_6$.
It leads to the group $O \simeq S_4$, the symmetry group of the
octahedron and the cube.
As it includes $T$ as a subgroup, the most general $O$-symmetric potential is $V_0$ from (\ref{V0Z2Z2Z3})
plus $V_T$ from (\ref{VT}) with the additional condition that $\tilde \lambda$ is real 
(the extra symmetry with respect to the $T$-symmetric case is a transposition of any two doublets).

\subsubsection{Including antiunitary transformations}

The case of $D_8$ has been already considered in section~\ref{subsubsection-D8Z2}. 

The tetrahedral potential $V_T + V_0$ from (\ref{VT}) and (\ref{V0Z2Z2Z3}) is symmetric under the following
antiunitary transformation:
\be
J' =  \left(\begin{array}{ccc}
0 & 1 &0 \\
1 & 0 & 0 \\
0 & 0 & -1
\end{array}\right)\cdot J\,,
\ee
which generates a $\Z_2^*$ group. Therefore the symmetry group of this potential is 
the full achiral tetrahedral group $T_d \simeq T \rtimes \Z_2^*$, which is isomorphic to $S_4$.

The octahedral potential is a particular case of the tetrahedral one,
therefore it is also invariant under an antiunitary transformation.
The extra $\Z_2^*$ subgroup is generated by the complex conjugation, $J$,
and this transformation commutes with the entire Higgs-family group $O$.
Therefore, the symmetry group of the potential is the full achiral octahedral 
symmetry group $O_h \simeq O \times \Z_2^*$.

\subsection{Extensions of abelian groups by an antiunitary transformation}\label{subsection-antiunitary-extensions}

The last type of extension we need to consider is of the type $A\arbitraryext \Z_2^*$,
where $A$ is one of the four abelian groups of Higgs-family transformations lying in a maximal torus,
that is, the first four groups in the list (\ref{abelian}), while the $\Z_2^*$ is as usual generated by an antiunitary transformation 
$J' = c J$. This problem was partly solved in \cite{abelianNHDM}, where such extensions leading to {\em abelian} groups
were analyzed. It was established that only the following four abelian groups of this type are realizable: 
$\Z_2^*$,  $\Z_4^*$, $\Z_2\times \Z_2^*$, and $\Z_2\times \Z_2\times \Z_2^*$.
Here, we consider {\em non-abelian} extensions of this type.

\subsubsection{Anti-unitary extension of $\Z_3$} 
The smallest non-abelian group we can have is $\Z_3 \rtimes \Z_2^* \simeq D_6$. We stress that this $D_6$ group we search for
is different from what we analyzed in section~\ref{subsection-D6}, because there the $D_6$ group contained only unitary transformations,
see a discussion in section~\ref{subsection-twoD6}.
Using the same notation for the generator $a$ of the $\Z_3$ group, we find that the transformation $c$ in the definition of $J'$
must be diagonal: $c = \mathrm{diag}(e^{i\xi_1}, e^{i\xi_2}, e^{-i(\xi_1+\xi_2)})$.
Then, studying how the $\Z_3$-symmetric potential $V_0+V_{\Z_3}$ changes under $J'=cJ$, 
we obtain that the only condition to be satisfied is (\ref{conditionD6Z2}).

If this condition is satisfied, then the potential is invariant under $\Z_3 \rtimes \Z_2^* \simeq D_6$,
if not, then the symmetry group remains $\Z_3$. This proves that both groups are realizable in 3HDM.
Note that in contrast with the $D_6 \times \Z_2^*$ case,
we do not place any extra condition such as (\ref{conditionV0}).

\subsubsection{Anti-unitary extension of $\Z_4$} 
A priori, the two non-abelian extensions here are again $D_8$ and $Q_8$.
With the usual convention for $a$, the generator of $\Z_4$, we again obtain that $c$ must be of the same diagonal form.
This immediately excludes the $Q_8$ case because we have $(J')^2 = c^*c = 1$.

The case of $\Z_4 \rtimes \Z_2^* \simeq D_8$ is possible. Even more, it turns out that the $\Z_4$-symmetric potential 
$V_0+V_{\Z_4}$ is {\em always} symmetric under some $J'$ of this type. It means, therefore, that if anti-unitary transformations
are included, $\Z_4$ is not realizable anymore: the true symmetry group of the potential is $\Z_4 \rtimes \Z_2^* \simeq D_8$.
In more physical terms, we conclude that {\em presence of a $\Z_4$ group of Higgs-family transformations makes the potential
explicitly $CP$-conserving}.

\subsubsection{Anti-unitary extension of $\Z_2 \times \Z_2$} 
The only non-abelian extension of the type $(\Z_2\times \Z_2)\arbitraryext \Z_2^*$ can produce only $D_8$, which was already considered.
We only remark here that $c$ turns out to be of the type (\ref{cgamma}), which places extra constraints on $V_0$.
Not satisfying these constraints will keep the symmetry group $\Z_2\times \Z_2$, which means that it is realizable.

\section{The $\Z_3 \times \Z_3$ chain}\label{sectionZ3Z3}

\subsection{The group and its extensions}\label{subsection-Z3Z3group}

The last abelian group from the list (\ref{abelian}), $\Z_3\times \Z_3$, requires a special treatment due to a number of reasons.
First, it does not belong to any maximal torus of $PSU(3)$ but is a maximal abelian subgroup of $PSU(3)$ on its own, \cite{abelianNHDM},
and its full preimage in $SU(3)$ is the non-abelian group $\Delta(27)$, \cite{fairbairn}.
Second, its automorphism group $Aut(\Z_3\times \Z_3)$ is sufficiently large and requires an accurate description.

Let us first remind how this group is constructed.
We first consider the subgroup of $SU(3)$ generated by
\be
a = \left(\begin{array}{ccc} 1 & 0 & 0 \\ 0 & \omega & 0 \\ 0 & 0 & \omega^2 \end{array}\right),\quad
b = \left(\begin{array}{ccc} 0 & 1 & 0 \\ 0 & 0 & 1 \\ 1 & 0 & 0 \end{array}\right)\,.
\ee
This group known as $\Delta(27)$ is non-abelian because $a$ and $b$ do not commute, but their commutator lies in the 
center of $SU(3)$:
\be
[a,b]=aba^{-1}b^{-1} = z^2 \in Z(SU(3))\,.
\ee
Therefore, its image under the canonical homomorphism
$SU(3) \to PSU(3)$ becomes the desired abelian group $\Delta(27)/\Z_3 = \Z_3 \times \Z_3$.
The true generators of $\Z_3 \times \Z_3$ are cosets $\bar a = aZ(SU(3))$ and $\bar b = bZ(SU(3))$ from $PSU(3)$,
and they obviously commute: $[\bar a,\bar b] = 1$.
Note that since $\Z_3 \times \Z_3$ is a maximal abelian subgroup in $PSU(3)$, there is no other element in $PSU(3)$
commuting with all elements of this group,
so $C_{PSU(3)}(\Z_3 \times \Z_3) = \Z_3 \times \Z_3$.

If the normal self-centralizing abelian subgroup of $G$, whose existence was proved in section~\ref{subsection-self-centralizing}, is $A=\Z_3\times \Z_3$,
then $G$ can be constructed as an extension of $A$ by a subgroup of $Aut(\Z_3\times \Z_3) = GL_2(3)$,
the general linear group of transformations of two-dimensional vector space over the finite field $\F_3$.
The order of this group is $|GL_2(3)|=48$, and it will prove useful if we now digress and describe the
structure of this group in some detail.

\subsubsection{$\Z_3 \times \Z_3$ as a vector space over $\F_3$}

The finite field $\F_3$ is defined as the additive group of integers $\mathrm{mod}\, 3$,
in which the multiplication is also introduced. It is convenient to denote the elements of this field as $0, 1, -1$
with obvious addition and multiplication laws.
Unlike the integers themselves, $\F_3$ is closed under division by a non-zero number, the property that makes $\F_3$ a field.

A vector space over a finite field is defined just as over any ``usual'' field.
The group $\Z_3 \times \Z_3$ can be thought of as a 2D vector space over $\F_3$; its elements
are (with the additive notation for the group operation) $\bar x = q_a \bar a + q_b \bar b$, where $q_a, q_b  \in \F_3$,
and $\bar a, \bar b$ are, as before, the generators of the group $\Z_3 \times \Z_3$. In the multiplicative notation,
we write $\bar x = \bar a^{q_a} \bar b^{q_b}$.

It is possible to define an antisymmetric scalar product in this space.
For any $\bar x \in \Z_3 \times \Z_3$, take any element of its preimage, $x \in \Delta(27)$.
Then, for any two elements $\bar x, \bar y \in \Z_3 \times \Z_3$, construct the number $(\bar x,\bar y)$ as $[x,y] \in \F_3$.
This map is faithful: although we can select different $x$ for a given $\bar x$, all of them give the same $[x,y]$.

Clearly, $(\bar x,\bar y) = - (\bar y,\bar x)$, in the additive notation.
Besides, the so defined product is linear in both arguments:
\be
(\bar x_1 + \bar x_2, \bar y) = (\bar x_1, \bar y) + (\bar x_2, \bar y)\,,
\quad
(\bar x, \bar y_1 + \bar y_2) = (\bar x, \bar y_1) + (\bar x, \bar y_2)\,.
\ee
Indeed, for any three elements of any group the following relation holds:
\be
[xy,z] = xyzy^{-1}x^{-1}z^{-1} = xyzy^{-1}\cdot z^{-1}x^{-1}xz\cdot x^{-1}z^{-1} =
x [y,z] x^{-1}[x,z]\,.
\ee
If in addition all commutators take values in the center of the group $SU(N)$, then $x$ and $x^{-1}$
can be cancelled, and we get $[xy,z]=[y,z][x,z]$.
In our case we represent $x_1 = \bar a^{q_{a1}} \bar b^{q_{b1}} z^{r_1}$ and similarly for $x_2$ and $y$,
and noting that all $z^{r_i}$ are inessential, we recover the above linearity in the first argument.
Thus, $\Z_3 \times \Z_3$ becomes a vector space over $\F_3$ equipped with an antisymmetric scalar product.

Note that all antisymmetric products in $\Z_3 \times \Z_3$ are proportional to $(\bar a,\bar b)$.
Indeed, if two elements $\bar x$ and $\bar x'$ are defined by their vectors
$\vec q = (q_{a},\,q_{b})$ and ${\vec q}' = (q'_{a},\,q'_{b})$,
then due to bilinearity we get
\be
(\bar x,\bar x') = (q_a q'_b - q_b q'_a) (\bar a,\bar b) = \epsilon_{ij}q_i q'_j (\bar a,\bar b)\,,\label{antisymmetric}
\ee
where $\epsilon_{ij}$ is the standard antisymmetric tensor with $\epsilon_{12} = - \epsilon_{21} = 1$, $\epsilon_{11}=\epsilon_{22}=0$.

\subsubsection{The automorphism group of $\Z_3 \times \Z_3$}

The automorphism group of $\Z_3\times \Z_3$ can then be viewed as the group of non-degenerate matrices
with elements from $\F_3$ acting in this 2D space, which explains why $Aut(\Z_3\times \Z_3) = GL_2(3)$.
Each matrix $q$ can be defined by its action on the generators $\bar a$, $\bar b$:
$\bar a \mapsto q_{aa} \bar a + q_{ab} \bar b$, $\bar b \mapsto q_{ba} \bar a + q_{bb} \bar b$, and can therefore be written as
\be
q = \mmatrix{q_{aa}}{q_{ab}}{q_{ba}}{q_{bb}}\,, \quad \det q \not = 0\,.
\ee
The group operation in $GL_2(3)$ is just the matrix product.

Recall now that the elements of both the $\Z_3 \times \Z_3$ group and of its automorphism group
are represented in our case as unitary or antiunitary transformations of the three doublets (that is,
we work not with the abstract groups but with their 3-dimensional complex representations).
Since $\Z_3 \times \Z_3$ is assumed to be normal in $G$, the elements $g \in Aut(\Z_3\times\Z_3)$
act on the elements of $\Z_3 \times \Z_3$ by conjugation:
$\bar x \mapsto g^{-1}\bar x g$, which we denoted by $g(\bar x)$.
Then the antisymmetric product defined above changes upon this action in the following way:
\be
(g(\bar x),g(\bar y)) = g^{-1}[x,y]g = g^{-1}z^r g =
\left\{\begin{array}{ll}
z^r = (\bar x,\bar y)\,, & \mbox{if $g$ is unitary}\,,\\
(z^*)^r = (z^{-1})^r = -(\bar x,\bar y)\,, & \mbox{if $g$ is anti-unitary}\,.\end{array}\right.
\ee
Here we used the fact the commutator of any two elements of $\Delta(27)$ lies in the center $Z(SU(3))$,
and that the $CP$ conjugation operator $J$ acts on any $x \in SU(3)$ by $J^{-1}xJ = x^*$.
So, unitary transformations preserve the antisymmetric product, while anti-unitary ones flip its sign.

Generically, the subgroup of a general linear group which conserves an antisymmetric bilinear
product in a vector space is called symplectic. Here we have the group $Sp_2(3)  <  GL_2(3)$.
It turns out that $Sp_2(3) = SL_2(3)$. Indeed, suppose $g \in GL_2(3)$ acts in the 2D space over $\F_3$
by mapping $q_i \mapsto g(q) = g_{ii'} q_{i'}$.
Then, the product transforms as
\be
(\bar x,\bar y) \mapsto (g(\bar x),g(\bar y)) = \epsilon_{ij}g_{ii'} g_{jj'}q^{(x)}_{i'} q^{(y)}_{j'} (\bar a,\bar b)
= \det g\cdot  (\bar x,\bar y)\,.
\ee
Since $\det g = \pm 1$, we get two kinds of transformations: those which conserve all products ($\det g = 1$, so that $g \in SL_2(3)$)
and those which flip their signs ($\det g = -1$), hence the identification of $Sp_2(3)$ and $SL_2(3)$ follows.

We conclude that the finite symmetry group $G$ of unitary transformations with the normal self-centralizing abelian subgroup $\Z_3\times \Z_3$
can be constructed as extension $(\Z_3\times \Z_3)\arbitraryext K$, where $K \leq SL_2(3)$.

\subsubsection{Explicit description of $SL_2(3)$}\label{listofSL23elemets}

The structure of the group $SL_2(3)$ is well-known, but it will prove useful to have the explicit
expressions for some of its elements.

The order of the group is $|SL_2(3)| =24$. It contains elements of order 2, 3, 4, and 6, generating the corresponding cyclic subgroups.
The subgroup $\Z_2$ is generated by the center of the group
\be
c = \mmatrix{-1}{0}{0}{-1}\,,
\ee
which in the multiplicative notation means $\bar a\mapsto \bar a^2$, $\bar b \mapsto \bar b^2$.
There are four distinct $\Z_3$ subgroups generated by
\be
f_1 = \mmatrix{1}{1}{0}{1}\,,\quad
f_2 = \mmatrix{1}{0}{1}{1}\,,\quad
f_3 = \mmatrix{0}{1}{-1}{-1}\,,\quad
f_4 = \mmatrix{0}{-1}{1}{-1}\,,\label{f1f4}
\ee
three $\Z_4$ subgroups generated by
\be
d_1 = \mmatrix{0}{1}{-1}{0}\,,\quad
d_2 = \mmatrix{1}{1}{1}{-1}\,,\quad
d_3 = \mmatrix{-1}{1}{1}{1}\,,\label{d1d2d3}
\ee
and four $\Z_6$ subgroups, which we do not write explicitly because 
they are absent in the list (\ref{abelian}).

Every element of $SL_2(3)$ can be represented by a unique (up to center) $SU(3)$ matrix,
which can be found by explicitly solving the corresponding matrix equations
defining the action of this element.
For example, the transformation $c$ is defined by
\be
c(a) = c^{-1} a c = a^2\,, \quad c(b) = c^{-1} b c = b^2\,.
\ee
Rewriting these equations as $3 \times 3$ matrix equations $ac = ca^2$, $bc = cb^2$
and solving them explicitly, we find the matrix $c$:
\be
c = \left(\begin{array}{ccc} -1 & 0 & 0 \\ 0 & 0 & -1 \\ 0 & -1 & 0 \end{array}\right)\,,\label{c3D}
\ee

\subsection{Generic potential}

A generic potential symmetric under $\Z_3\times \Z_3$ is
\bea
V & = &  - m^2 \left[\phi_1^\dagger \phi_1+ \phi_2^\dagger \phi_2+\phi_3^\dagger \phi_3\right]
+ \lambda_0 \left[\phi_1^\dagger \phi_1+ \phi_2^\dagger \phi_2+\phi_3^\dagger \phi_3\right]^2 \nonumber\\
&&+ {\lambda_1 \over \sqrt{3}} \left[(\phi_1^\dagger \phi_1)^2+ (\phi_2^\dagger \phi_2)^2+(\phi_3^\dagger \phi_3)^2
- (\phi_1^\dagger \phi_1)(\phi_2^\dagger \phi_2) - (\phi_2^\dagger \phi_2)(\phi_3^\dagger \phi_3)
- (\phi_3^\dagger \phi_3)(\phi_1^\dagger \phi_1)\right]\nonumber\\
&&+ \lambda_2 \left(|\phi_1^\dagger \phi_2|^2 + |\phi_2^\dagger \phi_3|^2 + |\phi_3^\dagger \phi_1|^2\right) \nonumber\\
&&+ \left(\lambda_3 \left[(\phi_1^\dagger \phi_2)(\phi_1^\dagger \phi_3) + (\phi_2^\dagger \phi_3)(\phi_2^\dagger \phi_1) + (\phi_3^\dagger 
\phi_1)(\phi_3^\dagger \phi_2)\right]
+ h.c.\right)\label{VZ3Z3}
\eea
with real $m^2$, $\lambda_0$, $\lambda_1$, $\lambda_2$ and complex $\lambda_3$. All values here are generic.
This potential can be found by taking the potential symmetric under the $\Z_3$ group of phase rotations described above
and then requiring that it be invariant under the cyclic permutations on the doublets.
Written in the space of bilinears, the potential has the form
\bea
V &=& - \sqrt{3}m^2 r_0 + 3\lambda_0 r_0^2 + \sqrt{3}\lambda_1 (r_3^2 + r_8^2) + \lambda_2 (|r_{12}|^2 + |r_{45}|^2 + |r_{67}|^2)\nonumber\\
&& + \lambda_3 (r_{12}r_{45}^* + r_{67}r_{12}^* + r_{45}r_{67}^*) + \lambda_3^* (r_{12}^*r_{45} + r_{67}^*r_{12} + r_{45}^*r_{67})\nonumber\\
&=& - \sqrt{3}m^2 r_0 + 3\lambda_0 r_0^2 + \Lambda_{ij}r_i r_j\,.
\label{VZ3Z3adj}
\eea
It is important to prove that this potential has no continuous symmetry.
Using the approach described in section~\ref{section-continuous}, we calculate the eigenvalues of $\Lambda_{ij}$
and find that it has four distinct eigenvalues of multiplicity two:
\be
\sqrt{3}\lambda_{1}\,,\quad  \lambda_2 + \lambda_3 + \lambda_3^*\,,\quad \lambda_2 + \omega \lambda_3 + \omega^2 \lambda_3^*\,,\quad
\lambda_2 + \omega^2 \lambda_3 + \omega \lambda_3^*\,.\label{spectrumLambda}
\ee
The first eigenvalue corresponds to the subspace $(r_3,r_8)$, while the rest are three 2D subspaces within its orthogonal complement
$(r_1,r_2,r_4,r_5,r_6,r_7)$.
For generic values of the coefficients, they do not coincide.
Then, according to our discussion in section~\ref{section-continuous},
a continuous symmetry group, if present, must consist only of phase rotations of the doublets.
But the $\lambda_3$ term selects only the $\Z_3$ group of phase rotations, 
which proves that no continuous symmetry leaves this potential invariant.

\subsection{Extension $(\Z_3\times \Z_3)\rtimes \Z_2$}

It turns out that $\Z_3\times \Z_3$ is not realizable because the potential (\ref{VZ3Z3}) is symmetric under a larger group
$(\Z_3\times \Z_3)\rtimes \Z_2 = \Delta(54)/\Z_3$, which is generated by $\bar a,\bar b,\bar c$ with the following relations
\be
\bar a^3 = \bar b^3 = 1,\ \bar c^2 = 1,\ [\bar a,\bar b]=1,\ \bar c\bar a\bar c = \bar a^2,\ \bar c\bar b\bar c = \bar b^2\,.\nonumber
\ee
In terms of explicit transformation laws, $\bar c$ is the coset $cZ(SU(3))$, with $c$ being the exchange of any two doublets,
for example (\ref{c3D}). Note that $\langle \bar a, \bar c\rangle = S_3$ is the group of arbitrary permutations of the three doublets.
Thus, if $G = (\Z_3\times \Z_3)\arbitraryext K$, then a $G$-symmetric potential must be a restriction of (\ref{VZ3Z3}),
and $K$ must contain a $\Z_2$ subgroup.

There are three kinds of subgroups of $SL_2(3)$ containing $\Z_2$ but not containing $\Z_6$:
$\Z_2$, $\Z_4$, and $Q_8$. In each case it would give a split extension, so $G$ must contain a subgroup
isomorphic to one of these groups.
Since, as we argued above, the quaternion group $Q_8$ is not realizable in 3HDM,
$K$ can only be $\Z_2$ or $\Z_4$. Therefore, the only additional case to consider is $(\Z_3\times \Z_3)\rtimes \Z_4$,
the group also known as $\Sigma(36)$, \cite{fairbairn}.

\subsection{Extension $(\Z_3\times \Z_3)\rtimes \Z_4$}

There are three distinct $\Z_4$ subgroups in $SL_2(3)$ generated by $d_1$, $d_2$, and $d_3$,
listed in (\ref{d1d2d3}). In principle, all of them are conjugate inside $SL_2(3)$, but for our purposes
all of them need to be checked.
Explicit solutions of the matrix equations give the following transformations:
\be
d_1 = {i \over\sqrt{3}} \left(\begin{array}{ccc} 1 & 1 & 1 \\ 1 & \omega^2 & \omega \\ 1 & \omega & \omega^2 \end{array}\right)\,,\quad
d_2 = {i \over\sqrt{3}} \left(\begin{array}{ccc} 1 & 1 & \omega \\ \omega & 1 & 1 \\ \omega & \omega^2 & \omega \end{array}\right)\,,\quad
d_3 = {i \over\sqrt{3}} \left(\begin{array}{ccc} 1 & 1 & \omega^2 \\ \omega & 1 & \omega \\ 1 & \omega & \omega 
\end{array}\right)\,.\label{d123}
\ee
Note that the prefactor $i/\sqrt{3}$ can also be written as $1/(\omega^2-\omega)$.

Let us mention here that when searching for explicit $SU(3)$ realizations of the transformations $d_1$,
we solve equations $d^{-1}_1ad_1=b$, $d_1^{-1}bd_1=a^2$. However, we could also use other representative matrices, $a'$ and $b'$,
which differ from $a$ and $b$ by transformations from the center. For example, we can also
ask for solutions of
\be
d_1^{\prime -1} a d'_1 = z^{n_1} b\,,\quad d_1^{\prime -1} b d'_1 = z^{n_2} a^2\,.
\ee
However, the solution of this equation can be written as
\be
d_1' = d_1 a^{n_1} b^{n_2}\,.
\ee
Therefore the resulting group $\langle \bar d'_1, \bar a, \bar b\rangle$ coincides with $\langle \bar d_1, \bar a, \bar b\rangle$.
The similar results hold for $d_2$ and $d_3$.

\subsubsection{Conditions for the $(\Z_3\times \Z_3)\rtimes \Z_4$ symmetry}

We should now check how the potential (\ref{VZ3Z3}) changes under these transformations and when it remains
invariant.
The calculation is simplified if we introduce the following combinations of bilinears (here $i^*j$ stands for $\phi_i^\dagger \phi_j$):
\bea
&&A_0 = 1^*1 + 2^*2 + 3^* 3\,,\quad A_1 = 1^*1 + \omega 2^*2 + \omega^2 3^* 3\,,\quad A_2 = A_1^*\nonumber\\
&&B_0 = 1^*2 + 2^*3 + 3^* 1\,,\quad B_1 = 1^*2 + \omega 2^*3 + \omega^2 3^* 1\,, \quad B_2 = 1^*2 + \omega^2 2^*3 + \omega 3^* 1\,,\nonumber\\
&&B_0^* = 2^*1 + 3^*2 + 1^* 3\,,\quad B_1^* = 2^*1 + \omega^2 3^*2 + \omega 1^* 3\,, \quad B_2^* = 2^*1 + \omega 3^*2 + \omega^2 1^* 
3\,.\nonumber
\eea
Next, introducing
\bea
X &=& {1 \over\sqrt{3}} \left[(\phi_1^\dagger \phi_1)^2+ (\phi_2^\dagger \phi_2)^2+(\phi_3^\dagger \phi_3)^2
- (\phi_1^\dagger \phi_1)(\phi_2^\dagger \phi_2) - (\phi_2^\dagger \phi_2)(\phi_3^\dagger \phi_3)
- (\phi_3^\dagger \phi_3)(\phi_1^\dagger \phi_1)\right]\,,\nonumber\\
&=& {1 \over \sqrt{3}}|A_1|^2\,,\nonumber\\
Y &=& |\phi_1^\dagger \phi_2|^2 + |\phi_2^\dagger \phi_3|^2 + |\phi_3^\dagger \phi_1|^2 = {|B_0|^2 + |B_1|^2 + |B_2|^2\over 3} \,,\nonumber\\
Z^* & = & (\phi_1^\dagger \phi_2)(\phi_1^\dagger \phi_3) + (\phi_2^\dagger \phi_3)(\phi_2^\dagger \phi_1)
+ (\phi_3^\dagger \phi_1)(\phi_3^\dagger \phi_2) = {|B_0|^2 + \omega^2 |B_1|^2 + \omega |B_2|^2\over 3}\,,
\eea
we write the potential (\ref{VZ3Z3}) as
\be
V = - \sqrt{3}m^2 r_0 + 3\lambda_0 r_0^2 + \lambda_i^* X_i\,,\quad \mbox{where}\quad
\lambda_i^* X_i = \lambda_1 X + \lambda_2 Y + \lambda_3 Z^* + \lambda_3^* Z
\ee
is the scalar product of the vector of coefficients and the vector of coordinates.
Now, it follows from explicit calculations that the action of $d_i$ can be compactly represented by the following transformations:
\bea
d_1: && A_1 \to B_0\,,\quad B_0 \to A_1^*\,, \quad B_1 \to \omega^2 B_2\,,\quad B_2 \to B_1^*\,,\nonumber\\
d_2: && A_1 \to B_1\,,\quad B_1 \to \omega A_1^*\,, \quad B_0 \to \omega B_2^*\,,\quad B_2 \to B_0\,,\nonumber\\
d_3: && A_1 \to B_2\,,\quad B_2 \to \omega A_1^*\,, \quad B_0 \to B_1\,,\quad B_1 \to \omega B_0^*\,,\nonumber
\eea
or even more compactly
\bea
d_1: && |A_1|^2 \leftrightarrow |B_0|^2\,,\quad |B_1|^2 \leftrightarrow |B_2|^2\,,\nonumber\\
d_2: && |A_1|^2 \leftrightarrow |B_1|^2\,,\quad |B_0|^2 \leftrightarrow |B_2|^2\,,\nonumber\\
d_3: && |A_1|^2 \leftrightarrow |B_2|^2\,,\quad |B_0|^2 \leftrightarrow |B_1|^2\,.
\eea
Therefore, their action in the space of $(X, Y, Z, Z^*)$ is given by the following hermitean and
unitary matrices
\be
T(d_1) = {1 \over 3}\left(\!\begin{array}{cccc} 0 & \sqrt{3} & \sqrt{3} & \sqrt{3} \\
\sqrt{3} & 2 & -1 & -1 \\
\sqrt{3} & -1 & -1 & 2 \\
\sqrt{3} & -1 & 2 & -1 \end{array}\!\right)\,,\quad
T(d_2) = {1 \over 3}\left(\!\begin{array}{cccc} 0 & \sqrt{3} & \omega^2\sqrt{3} & \omega\sqrt{3} \\
\sqrt{3} & 2 & -\omega^2 & -\omega \\
\omega\sqrt{3} & -\omega & -1 & 2\omega^2 \\
\omega^2 \sqrt{3} & -\omega^2 & 2\omega & -1 \end{array}\!\right)\,,\nonumber
\ee
and $T(d_3) = [T(d_2)]^*$.
It can be also noted that $T(d_2)$ acts in the space of $(X,\,Y,\,\omega^2 Z,\, \omega Z^*)$ by the matrix $T(d_1)$.
So, $T(d_1)$, $T(d_2)$ and $T(d_3)$ represent the same type of transformations acting in the spaces
$(X,\,Y,\, Z,\, Z^*)$, $(X,\,Y,\,\omega^2 Z,\, \omega Z^*)$, or $(X,\,Y,\,\omega Z,\, \omega^2 Z^*)$, respectively.
That is, if $(x,y,z,z^*)$ is an eigenvector of $T(d_1)$, then $(x,y,\omega z, \omega^2 z^*)$ is an eigenvector of $T(d_2)$
and $(x,y,\omega^2 z, \omega z^*)$ is an eigenvector of $T(d_3)$.
This observation restores the expected symmetry among the three types of $\Z_4$ subgroups inside $SL_2(3)$.

Since these matrices are hermitean and unitary, they act by pure reflections,
which implies that each of them is diagonalizable and has eigenvalues $\pm 1$.
If we want the potential to be symmetric under one of these $d_i$,
it must induce the same transformations in the space of $\lambda_i = (\lambda_1, \lambda_2,\lambda_3^*,\lambda_3)$.
Therefore, in order to find conditions that the potential is invariant under $d_i$,
we need to find eigenvectors of $T(d_i)$ corresponding to the eigenvalue $-1$ and require that $\lambda_i$'s
projection on these eigenvectors is zero.

Consider first $T(d_1)$. It has two eigenvectors corresponding to the eigenvalue $-1$:
$(-\sqrt{3},1,1,1)$ and $(0,0,1,-1)$. Therefore, we obtain the following condition for the potential to be symmetric under $d_1$:
\be
\lambda_3 \ \mbox{is real and }\lambda_3 = {\sqrt{3}\lambda_1 - \lambda_2 \over 2}\,.
\ee
Similarly, for $d_2$ we have
\be
\omega \lambda_3 \ \mbox{is real and }\omega \lambda_3 = {\sqrt{3}\lambda_1 - \lambda_2 \over 2}\,.
\ee
For $d_3$ we have the complex conjugate condition.
Therefore, the potential (\ref{VZ3Z3}) is symmetric under $(\Z_3\times\Z_3)\rtimes \Z_4$
if 
\be
\left({2 \lambda_3 \over \sqrt{3}\lambda_1 - \lambda_2}\right)^3 = 1\,,\label{conditionZ3Z3Z4}
\ee
which encompasses all these cases.
Let us also mention that when these conditions are taken into account, the spectrum of the matrix $\Lambda_{ij}$  given
in (\ref{spectrumLambda})
becomes even more degenerate: it contains two eigenvalues of multiplicity four (we refer to this spectrum as $4+4$).

\subsubsection{Absence of a continuous symmetry}

In order for the group $(\Z_3\times \Z_3)\rtimes \Z_4$ to be realizable, we need to show that
the potential (\ref{VZ3Z3}) with parameters satisfying (\ref{conditionZ3Z3Z4}) is not symmetric
under any continuous group.

We first note that even if such a continuous symmetry group existed, it could only be $U(1)$. 
Indeed, the spectrum of $\Lambda_{ij}$ in our case
is $4+4$, while for $U(1)\times U(1)$ and $SU(2)$ it must be $6+2$,
and for $SO(3)$ it must be $5+3$.

Let us now consider, for example, the $d_1$-symmetric potential.
Using $\sum_{i=1}^8 r_i^2 = \alpha r_0^2$, where $1/4 \le \alpha \le 1$ parametrizes $SU(3)$-orbits
in the orbit space,
we can rewrite it as
\be
V = - \sqrt{3}m^2 r_0 + (3\lambda_0 +\alpha\sqrt{3}\lambda_1)r_0^2 - {\sqrt{3}\lambda_1-\lambda_2\over 2}
\left(|r_{12}-r_{45}|^2 + |r_{45}-r_{67}|^2 + |r_{67}-r_{12}|^2\right)\,. \label{Vn}
\ee
Suppose the potential (\ref{Vn}) is invariant under a $U(1)$ group of transformations of doublets,
generated by the generator $t$ from the algebra $su(3)$. Since the potential (\ref{Vn}) is invariant under the $S_3$
group of
arbitrary permutations of the doublets, then the same potential must be also invariant under other $U(1)$ subgroups
which are generated by various $t^g$, which are obtained by acting on $t$ by $g \in S_3$.
If $t \not = t^g$ (or to be more accurate, if their corresponding $U(1)$ groups are different),
then the continuous symmetry group immediately becomes larger than $U(1)$, which is impossible.
Therefore, $t^g$ must be equal (up to sign) to $t$ for all $g \in S_3$.
In other words, $S_3$ must stabilize the $U(1)$ symmetry group.

There exist only two elements in the algebra $su(3)$ with this property:
\be
t_1 = \left(\begin{array}{ccc}
0 & i & -i \\
-i & 0 & i \\
i & -i & 0
\end{array}\right)\quad\mbox{and}\quad
t_2 = \left(\begin{array}{ccc}
0 & 1 & 1 \\
1 & 0 & 1 \\
1 & 1 & 0
\end{array}\right)\,.
\ee
$t_2$ generates pure phase rotations.
It is explicitly $S_3$-invariant, therefore the corresponding $U(1)$ group is also invariant.
$t_1$ induces $SO(3)$ rotations of the doublets around the axis $(1,1,1)$.
It is $\Z_3$-invariant, while reflections from $S_3$
flip the sign of $t_1$. However the $U(1)$ group is still invariant.
Since $t_1$ and $t_2$ realize different representations of $S_3$, one cannot
take their linear combinations. So, the list of possibilities is restricted only to $t_1$ and $t_2$ themselves.

The eigenvalues and eigenvectors of $t_1$ are
\be
\zeta=0:\ \stolb{1}{1}{1}\,,\quad
\zeta = \sqrt{3}:\ \stolb{1}{\omega^2}{\omega}\,,\quad
\zeta = -\sqrt{3}:\ \stolb{1}{\omega}{\omega^2}\,.
\ee
The presence of the eigenvalue $\zeta=0$ implies that the combination $\phi_1 + \phi_2 + \phi_3$ is invariant under the $U(1)$ group generated by $t_1$.
Bilinear invariants are 
\be
|\phi_1 + \phi_2 + \phi_3|^2\,,\quad
|\phi_1 + \omega^2 \phi_2 + \omega \phi_3|^2\,,\quad
|\phi_1 + \omega \phi_2 + \omega^2 \phi_3|^2\,,\quad
\ee
which simply means that $r_1 + r_4 + r_6$ and $r_2 + r_5 + r_7$ are, separately, invariant.
So, if the potential depends only on $r_0$ and these two combinations, then it is symmetric under the $U(1)$ generated by $t_1$.
The point is that our potential (\ref{Vn}) cannot be written via these combinations only,
therefore it is not invariant under this group.

Consider now $t_2$. Its eigensystem is
\be
\zeta=2:\ \stolb{1}{1}{1}\,,\quad
\zeta = -1:\ \stolb{0}{1}{-1}\ \mbox{and}\ \stolb{2}{-1}{-1}\,.
\ee
There is no zero eigenvalue, therefore no linear combination of $\phi$'s is invariant.
The independent bilinear combinations are
\be
|\phi_1 + \phi_2 + \phi_3|^2\,,\quad
|\phi_2 - \phi_3|^2\,,\quad |2\phi_1 -\phi_2 - \phi_3|^2\,,\quad
(\phi_2^\dagger - \phi_3^\dagger)(2\phi_1 -\phi_2 - \phi_3)\,.
\ee
In addition, there exists a triple product of $\phi$'s which is also invariant but it is irrelevant for our analysis because
our potential contains only two $\phi$'s and two $\phi^\dagger$'s.
These invariants can also be rewritten as the following linearly independent invariants (here $\rho_i = \phi_i^\dagger \phi_i$):
\be
\rho_1 + 2r_6\,,\quad \rho_2 + 2r_4\,,\quad \rho_3 + 2 r_1\,,\quad r_2 + r_5 + r_7\,.
\ee
Despite the fact that we now have more invariants than in the previous case,
it is still impossible to express (\ref{Vn}) via these combinations.
This means that (\ref{Vn}) is not symmetric under $t_2$.

This completes the proof that the potential (\ref{VZ3Z3}) subject to conditions (\ref{conditionZ3Z3Z4})
is not invariant under any continuous group.

\subsubsection{Absence of a larger finite symmetry group}

Although the group-theoretic arguments guarantee that no other extension can be used,
it is still instructive to check what happens if we try to impose invariance under other subgroups of $SL_2(3)$.

Let us first note that if we try to impose simultaneous invariance under two among $d_i$ (trying to get $Q_8$),
we must set $\lambda_3 = 0$. But then the potential has an obvious continuous symmetry, and our attempt fails.

Next, let us assume that the potential is invariant under $(\Z_3\times \Z_3)\rtimes \Z_3$, where the last $\Z_3$ is generated
by one of the generators $f$ in (\ref{f1f4}), for example $f=f_1$. Its representative matrix in $SU(3)$ is
\be
f = {-i \over\sqrt{3}} \left(\begin{array}{ccc} 1 & \omega^2 & 1 \\ 1 & 1 & \omega^2 \\ \omega^2 & 1 & 1  \end{array}\right)\,,\quad
f^3 = 1\,.
\ee
An analysis similar to what was described above allows us to find the corresponding transformation matrix
in the space of $X, Y, Z, Z^*$:
\be
T(f_1) = {1 \over 3}\left(\begin{array}{cccc} 0 & \sqrt{3} & \sqrt{3}\omega^2 & \sqrt{3}\omega  \\
\sqrt{3} & 2 & -\omega^2 & -\omega \\
\sqrt{3}\omega^2 & -\omega^2 & -\omega & 2 \\
\sqrt{3}\omega & -\omega & 2 & -\omega^2 \end{array}\right)\,.
\ee
It leads to the following conditions for the potential to be symmetric under $(\Z_3\times \Z_3)\rtimes \Z_3$:
\be
\lambda_3 = \lambda_3^*\quad\mbox{and}\quad \lambda_{1} = {\lambda_2 - \lambda_3\over \sqrt{3}}\,.
\label{conditionZ3Z3Z3}
\ee
In the space of bilinears, the potential can then be compactly written as
\be
V = - \sqrt{3}m^2 r_0 + (3\lambda_0 +\sqrt{3}\lambda_1\alpha)r_0^2 + \lambda_3  |r_{12}+ r_{45} + r_{67}|^2\,.
\label{VZ3Z3Z6adj}
\ee
The spectrum of $\Lambda_{ij}$ becomes of the type $6+2$. This high symmetry hints at existence
of a possible continuous symmetry of the potential, and it is indeed the case.
For example, the following $SO(2)$ rotations among three doublets, $\phi_a \mapsto R_{ab}(\alpha)\phi_b$,
leave $r_{12}+ r_{45} + r_{67}$ invariant:
\be
\label{z3continuous}
R(\alpha) =
{1\over 3}\left(\begin{array}{ccc}
1+2\cos\alpha & 1+2\cos\alpha'' & 1+2\cos\alpha' \\
1+2\cos\alpha' & 1+2\cos\alpha & 1+2\cos\alpha'' \\
1+2\cos\alpha'' & 1+2\cos\alpha' & 1+2\cos\alpha
\end{array}\right)\,,
\ee
with $\alpha \in [0,2\pi)$ and $\alpha' = \alpha + 2\pi/3$ and $\alpha'' = \alpha + 4\pi/3$.
Note that at $\alpha = 0$, $2\pi/3$ and $4\pi/3$ we recover the $\Z_3$ group $\langle b\rangle$.

We conclude therefore that imposing invariance under $\Z_3 < SL_2(3)$ makes the potential symmetric
under a continuous group.
In this way, we completely exhausted possibilities offered by $SL_2(3)$.

\subsection{Anti-unitary transformations}

We showed in section~\ref{subsection-Z3Z3group}
that antiunitary transformations correspond to elements of $GL_2(3)$ not lying in $SL_2(3)$
as they have negative determinant and flip the sign of the antisymmetric scalar product in $A=\Z_3\times \Z_3$.
The complex conjugation operator, $J$, acts in $A$ by sending $a$ to $a^2$ and leaving $b$ invariant.
Therefore, the corresponding matrix is
\be
J = \mmatrix{-1}{0}{0}{1}\,.
\ee
Since any antiunitary transformation can be written as $J' = q J$,
where $q$ is unitary, it follows that $q$ must belong to $SL_2(3)$.

Next, we need to find which $q$'s can be used. Clearly, $(J')^2 = q J q J = qq^* \in SL_2(3)$.
If we are looking for an antiunitary symmetry of a $(\Z_3\times\Z_3)\rtimes \Z_2$-symmetric potential,
then $qq^*$ must be either $1$ or $c$, which generates the center of $SL_2(3)$.

Let us first consider the second possibility.
\be
\mbox{If}\ \ q = \mmatrix{x}{y}{z}{t}\,,\quad \mbox{then}\ \ q^* = \mmatrix{x}{-y}{-z}{t}\,.
\ee
Using this to solve $qq^* = c$, we get six possible solutions, but all of them have $\det q = -1$,
that is, they do not belong to $SL_2(3)$. Therefore, the only possibility is $qq^* = 1$.

But then we can apply the results of our search for antiunitary transformations for the $D_6$ case.
Our group $(\Z_3\times\Z_3)\rtimes \Z_2$ contains the $D_6$ subgroup with $\delta = \pi$.
Therefore, we arrive at the conclusion: in order for our potential to be symmetric under an antiunitary transformation,
we must require
\be
6 \arg \lambda_3 = 0\,.\label{6arg3}
\ee
If this criterion is satisfied, the symmetry group becomes $(\Z_3\times\Z_3)\rtimes (\Z_2\times \Z_2^*)$;
otherwise the group remains $(\Z_3\times\Z_3)\rtimes \Z_2$. Therefore, both groups are realizable in 3HDM.

Now, consider the case of the extended symmetry group, $(\Z_3\times\Z_3)\rtimes \Z_4 \simeq \Sigma(36)$.
In this case (\ref{6arg3}) is satisfied automatically due to (\ref{conditionZ3Z3Z4}).
We then conclude that in this case the realizable symmetry is $\Sigma(36)\rtimes \Z_2^*$.

\section{Summary and discussion}\label{section-summary}

\subsection{List of realizable finite symmetry groups in 3HDM}

Bringing together the results of the search for abelian symmetry groups \cite{abelianNHDM}
and of the present work, we can finally give the list of finite groups which can appear as the
symmetry groups of the scalar sector in 3HDM.
If only Higgs-family transformations are concerned, the realizable finite groups are
\bea
&&\Z_2, \quad \Z_3,\quad \Z_4,\quad \Z_2\times\Z_2,\quad D_6,\quad D_8, \quad T\simeq A_4,\quad O\simeq S_4\,, \nonumber\\
&& (\Z_3\times\Z_3)\rtimes \Z_2 \simeq \Delta(54)/\Z_3,\quad (\Z_3\times\Z_3)\rtimes \Z_4\simeq \Sigma(36)\,.
\label{groups1}
\eea
This list is complete: trying to impose any other finite symmetry group of Higgs-family transformations
leads to the potential with a continuous symmetry.

Fig.~\ref{fig-tree} should help visualize relations among different groups from this list.
Going up along a branch of this tree means that, starting with a potential symmetric under the lower group,
one can restrict its free parameters in such a way that the potential becomes symmetric under the upper group.
\tikzset{node distance=1.2cm, auto}
\begin{figure}[!htb]
   \centering
\begin{tikzpicture}
  \node (E) {$\{e\}$};
  \node (Z2) [above of=E] {$\Z_2$};
  \node (Z2Z2) [above of=Z2] {$\Z_2\times\Z_2$};
  \node (Z3) [node distance=2cm, right of=Z2Z2] {$\Z_3$};
  \node (Z4) [node distance=2cm, left of=Z2Z2] {$\Z_4$};
  \node (D8) [above of=Z4] {$D_8$};
  \node (A4) [above of=Z2Z2] {$A_4$};
  \node (S4) [above of=A4] {$S_4$};
  \node (D6)[node distance=2cm, right of=A4]{$D_6$};
  \node (D54) [above of=D6] {$\Delta(54)/\Z_3$};
  \node (S36) [above of=D54] {$\Sigma(36)$};
  \draw[->] (E) to node {} (Z2);
  \draw[->] (E) to node {} (Z3);
  \draw[->] (Z2) to node {} (Z4);
  \draw[->] (Z2) to node {} (Z2Z2);
  \draw[->] (Z2) to node {} (D6);
  \draw[->] (Z3) to node {} (D6);
  \draw[->] (Z3) to node {} (A4);
  \draw[->] (Z4) to node {} (D8);
  \draw[->] (Z2Z2) to node {} (D8);
  \draw[->] (Z2Z2) to node {} (A4);
  \draw[->] (A4) to node {} (S4);
  \draw[->] (D8) to node {} (S4);
  \draw[->] (D6) to node {} (S4);
  \draw[->] (D6) to node {} (D54);
  \draw[->] (D54) to node {} (S36);
\end{tikzpicture}
\caption{Tree of finite realizable groups of Higgs-family transformations in 3HDM}
   \label{fig-tree}
\end{figure}
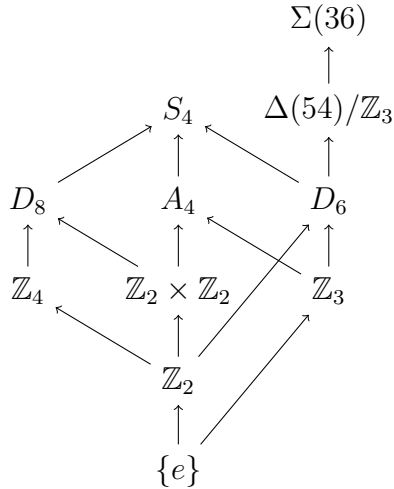

If both unitary (Higgs-family) and antiunitary (generalized-$CP$) transformations are allowed, the list becomes
\bea
&&\Z_2, \qquad \Z_3,\qquad \Z_2\times\Z_2, \qquad \Z_2^*, \qquad \Z_4^*,\nonumber\\
&&\Z_2\times\Z_2^*,\qquad \Z_2\times \Z_2\times\Z_2^*,\qquad \Z_3 \rtimes \Z_2^* \simeq D_6,\qquad \Z_4 \rtimes \Z_2^* \simeq D_8\,, \nonumber\\
&& D_6, \qquad D_6\times \Z_2^*, \qquad D_8\times \Z_2^*, \qquad A_4 \rtimes \Z_2^*\simeq T_d \,,\qquad S_4 \times \Z_2^*\simeq O_h\,,\nonumber\\
&&(\Z_3\times\Z_3)\rtimes \Z_2,\qquad (\Z_3\times\Z_3)\rtimes (\Z_2\times \Z_2^*),\qquad \Sigma(36)\rtimes \Z_2^*\,.
\label{groups2}
\eea
As usual, an asterisk here indicates that the generator of the corresponding group
is an anti-unitary transformation.
Note that Higgs-family transformation groups $\Z_4$, $D_8$, $A_4$, $S_4$, and $\Sigma(36)$ become non-realizable in this case,
because potentials symmetric under them are automatically symmetric under an additional anti-unitary transformation.
In all cases apart from $A_4$ this is a consequence of our finding in section~\ref{subsection-antiunitary-extensions} 
that presence of the $\Z_4$ group of Higgs-family transformations {\em always} leads to an additional anti-unitary symmetry.

These lists complete the classification of realizable finite symmetry groups of the scalar sector of 3HDM.
Conditions for the existence and examples of the potentials symmetric under each of these groups have been given
in \cite{abelianNHDM} and in the present work. For the reader's convenience, we collect examples 
with non-abelian groups in the Appendix.

\subsection{Interplay between Higgs-family symmetries and explicit $CP$-violation}

In 2HDM, presence of any Higgs-family symmetry immediately leads to
a generalized-$CP$ symmetry. In other words, it is impossible to write down an explicitly $CP$-violating
2HDM potential with {\em any} Higgs-family symmetry.
In this sense, generalized-$CP$ symmetries can be viewed as the smallest building blocks 
of any symmetry group in 2HDM.

By comparing lists (\ref{groups1}) and (\ref{groups2}), we see that this conclusion is no longer true for 3HDM,
namely there are some Higgs-family symmetry groups which are compatible with explicit $CP$-violation.
However we found another, quite remarkable feature in 3HDM: {\em the presence of a $\Z_4$ group of Higgs-family transformations
guarantees that the potential is explicitly $CP$-conserving}. This is, of course, a sufficient but not necessary condition for explicit $CP$-violation.
Put in other words, explicit $CP$-violation is incompatible with the Higgs-family symmetry group $\Z_4$. 

\subsection{Two different $D_6$ groups}\label{subsection-twoD6}

It is interesting to note that the list (\ref{groups2}) contains two different $D_6$ groups.
One is $\Z_3 \rtimes \Z_2^*$, generated by a Higgs-family transformation of order 3 and a generalized-$CP$ transformation.
The other $D_6$ is a group of Higgs-family transformations only, and a potential invariant under it
does not have any generalized-$CP$ symmetry. Clearly, they lead to different phenomenological consequences,
as the first case is explicitly $CP$-conserving, while the latter is explicitly $CP$-violating.

Such a situation was absent in the two-Higgs-doublet model, where fixing the symmetry group
uniquely defined the (tree-level) phenomenological consequences in the scalar sector.
What makes it possible in 3HDM is a looser relation between Higgs-family and generalized-$CP$ symmetries
just discussed. In particular, it is possible to have a potential with the Higgs-family $D_6$ symmetry group
without any generalized-$CP$ symmetry. 2HDM does not offer this kind of freedom: any non-trivial Higgs-family symmetry 
group automatically leads to a generalized-$CP$ symmetry.


\subsection{Further directions of research}

Certainly, our results do not provide answers to {\em all} symmetry-related questions which can be posed in 3HDM.
Our paper should rather be regarded as the first step towards systematic exploration of all the possibilities
offered by three Higgs doublets. Here are some further questions which deserve a closer study:
\begin{itemize}
\item
Continuous symmetry groups should also be included in the list.
There exist only few Lie groups inside $PSU(3)$: $U(1)$, $U(1)\times U(1)$, $SU(2)$, $SU(2)\times U(1)$,
$SO(3)$. The non-trivial question is which of these groups can be merged with some of the finite groups
and with anti-unitary transformations (the case of abelian groups was analyzed in \cite{abelianNHDM}).
\item
It is well-known that the vacuum state does not have to respect all the symmetries of the Lagrangian,
so the finite symmetry groups described here can be broken upon electroweak symmetry breaking.
What are the symmetry breaking patterns for each of these groups?
Clearly, if the symmetry group is very small, then the vacuum state can either conserve it or break it, either completely
or partially.
But when the finite group becomes sufficiently large, there are two important changes.
First, some of the groups can never be conserved upon EWSB; the origin of this feature and
some 3HDM examples were discussed in \cite{frustrated}.
Second, a sufficiently large symmetry group cannot break down completely,
as it would create too many
degenerate vacua, which is not possible from the algebraic-geometric point of view.
Indeed, in the geometric reformulation of the Higgs potential minimization problem \cite{ivanovNHDM2},
the points of the global minima in the $(r_0, r_i)$-space are precisely the contact points of two 9-dimensional algebraic manifolds:
the orbit space and a certain quadric. Intersection of two algebraic manifolds of known degrees is also an algebraic
manifold of a certain degree (the planar analogue of this statement is the Bezout's theorem). 
In the degenerate case when this manifold is reduced to a set of isolated points, 
there must exist an upper limit for the number of these points. Unfortunately, we have not yet found this number
for 3HDM, but its existence is beyond any doubt.
\item
What are possible symmetries of the potential beyond the unitary and antiunitary transformations?
For example, the full reparametrization group of the 2HDM potential is $GL(2,\CC) \rtimes \Z^*_2$ rather than $SU(2) \rtimes Z^*_2$,
\cite{ivanov2HDM}. It means that a potential can be left invariant by transformations
which are neither unitary nor anti-unitary. Although these transformations played important
role in the geometric constructions in the 2HDM orbit space,
they did not produce new symmetry groups beyond what was already found from the unitary transformations.
It would be interesting to check the situation in 3HDM. Unfortunately, the geometric
method which worked well for 2HDM becomes much more intricate with more than two doublets, \cite{ivanovNHDM,ivanovNHDM2}.
\item
It would also be interesting to see if the potential can have symmetries beyond re\-pa\-ra\-met\-ri\-za\-ti\-on transformations.
In the case of 2HDM, this problem was analyzed in \cite{pilaftsis}.
Although these additional symmetries cannot be extended to kinetic term,
they could still provide useful information on the structure of the Higgs potential and
properties of the physical Higgs bosons.
\end{itemize}

In summary, we found all finite groups which can be realized as symmetry groups of
Higgs-family or generalized-$CP$ transformations in the three-Higgs-doublet model.
Our list (\ref{groups2}) is complete: trying to impose any other discrete symmetry group
on the 3HDM Higgs potential will make it symmetric under a continuous group.\\

This work was supported by the Belgian Fund F.R.S.-FNRS,
and in part by grants RFBR 11-02-00242-a, RFBR 12-01-33102, RF President grant for
scientific schools NSc-3802.2012.2, and the
Program of Department of Physics SC RAS and SB RAS "Studies of Higgs boson and exotic particles at LHC."

\appendix

\section{3HDM potentials with non-abelian Higgs-family symmetry group}


Here, for the reader's convenience, we list once again Higgs potentials with a given symmetry group.
We focus here on cases with non-abelian groups from the list (\ref{groups2})
because abelian ones were already discussed in detail in \cite{abelianNHDM}.
In each case we start from the most general potential compatible with the given realizable group presented in the main text 
and use the residual reparametrization freedom to simplify the coefficients of the potential 
(usually, it amounts to rephasing of doublets which makes some of the coefficients real). 
For each group $G$, the potential written below faithfully represents all possible Higgs potentials 
with realizable symmetry group $G$. In this sense, the symmetry group uniquely defines the phenomenology
of the scalar sector of 3HDM, the only exception being $D_6$ with its two distinct realizations.

{\bf \boldmath Group $D_6 \simeq \Z_3 \rtimes \Z_2^*$.}
Consider the most general phase-independent part of the Higgs potential
\be
V_0 =  - \sum_{1\le i \le 3} m_i^2(\phi_i^\dagger \phi_i) + \sum_{1 \le i\le j \le 3} \lambda_{ij} (\phi_i^\dagger \phi_i)(\phi_j^\dagger \phi_j)\nonumber\\
+ \sum_{1 \le i < j \le 3} \lambda'_{ij} (\phi_i^\dagger \phi_j)(\phi_j^\dagger \phi_i)\,,\label{Tsymmetric2}
\ee
and the additional terms
\be
V_{\Z_3} = \lambda_1 (\phi_2^\dagger \phi_1)(\phi_3^\dagger \phi_1) + \lambda_2(\phi_1^\dagger \phi_2)(\phi_3^\dagger \phi_2) 
+ \lambda_3 (\phi_1^\dagger \phi_3)(\phi_2^\dagger \phi_3) + h.c.\label{VZ3-again}
\ee
For generic $\lambda_i$, these terms are symmetric only under the group $\Z_3$ generated by
\be
a_3 = \left(\begin{array}{ccc} \omega & 0 & 0 \\ 0 & \omega^2 & 0 \\ 0 & 0 & 1 \end{array}\right)\,,\quad \omega = \exp\left({2\pi i \over 3}\right)\,.
\label{generator-a3}
\ee
If it happens that the product $\lambda_1\lambda_2 \lambda_3$ is purely real, then by rephasing of doublets 
one can make all coefficients in (\ref{VZ3-again}) real. 
The resulting potential, $V_0 + V_{\Z_3}$,
is symmetric under $D_6 \simeq \Z_3 \rtimes \Z_2^*$ generated by $a_3$ and the $CP$-transformation.

{\bf \boldmath Group $D_8 \simeq \Z_4 \rtimes \Z_2^*$.} Consider now terms
\be
V_{\Z_4} = \lambda_1 (\phi_3^\dagger \phi_1)(\phi_3^\dagger \phi_2) + \lambda_2 (\phi_1^\dagger \phi_2)^2 + h.c., \label{VZ4-again}
\ee
which are symmetric under the group $\Z_4$ generated by
\be
a_4 = \left(\begin{array}{ccc} i & 0 & 0 \\ 0 & -i & 0 \\ 0 & 0 & 1 \end{array}\right)\,.\label{generator-a4}
\ee
It is always possible to compensate the phases of $\lambda_1$ and $\lambda_2$ by an appropriate rephasing of the doublets.
Therefore, the potential $V_0 + V_{\Z_4}$
is symmetric under the group $D_8 \simeq \Z_4 \rtimes \Z_2^*$ generated by $a_4$ and the $CP$-transformation.

{\bf \boldmath Group $D_6$ of unitary transformations.} Let us restrict the coefficients of $V_0$ in the way that guarantees
the symmetry under $\phi_1 \leftrightarrow \phi_2$. Then, $V_0$ turns into
\bea
V_{1} &=& - m_{11}^2\left[(\phi_1^\dagger \phi_1) + (\phi_2^\dagger \phi_2)\right]  - m_{33}^2 (\phi_3^\dagger \phi_3)
+ \lambda_{11}\left[(\phi_1^\dagger \phi_1)^2 + (\phi_2^\dagger \phi_2)^2\right]  + \lambda_{33}(\phi_3^\dagger \phi_3)^2\label{Vcommon}\\
&&\!\!\!\!+ \lambda_{13}\left[(\phi_1^\dagger \phi_1) + (\phi_2^\dagger \phi_2)\right](\phi_3^\dagger \phi_3) + \lambda_{12}(\phi_1^\dagger 
\phi_1)(\phi_2^\dagger \phi_2)
+ \lambda'_{13}\left[|\phi_1^\dagger \phi_3|^2 + |\phi_2^\dagger \phi_3|^2\right] + \lambda'_{12}|\phi_1^\dagger \phi_2|^2\,,
\nonumber
\eea
where all coefficients are real and generic.
Imposing the same requirement on $V_{\Z_3}$ and performing rephasing, we obtain
\be
V_{D_6} = \lambda_1\left[(\phi_2^\dagger \phi_1)(\phi_3^\dagger \phi_1) - (\phi_1^\dagger \phi_2)(\phi_3^\dagger \phi_2) \right]
+ |\lambda_3| e^{i\psi_3} (\phi_1^\dagger \phi_3)(\phi_2^\dagger \phi_3) + h.c.\label{VD6}
\ee
where $\lambda_1$ is real and $\sin\psi_3 \not = 0$.
The resulting potential, $V_1 + V_{D_6}$, is symmetric under $D_6$ generated by $a_3$ and 
\be
b = \left(\begin{array}{ccc} 0 & 1 & 0 \\ 1 & 0 & 0 \\ 0 & 0 & -1 \end{array}\right)\,.
\ee
There are no other Higgs-family or generalized-$CP$ transformations which leave this potential invariant.
Any explicitly $CP$-violating $D_6$-symmetric 3HDM potential can always be brought into this form.

{\bf \boldmath Group $D_6 \times \Z_2^*$.} If in the previous case we set $\sin\psi_3 = 0$ in (\ref{VD6}), then the potential
becomes symmetric under $D_6 \times \Z_2^*$ generated by $a_3$, $b$, and the generalized $CP$-transformation $b\cdot CP$.

{\bf \boldmath Group $D_8 \times \Z_2^*$.}
The potential $V_1 + V_{\Z_4}$ is symmetric under the group $D_8 \times \Z_2^*$ generated by $a_4$, $b$, and $b\cdot CP$.

{\bf \boldmath Group $A_4 \rtimes \Z_2^*$.}
A potential symmetric under $A_4 \rtimes \Z_2^*$ can be brought into the following form
\bea
V_{A_4 \rtimes \Z_2^*} &=& -m^2 \left[\phi_1^\dagger\phi_1 + \phi_2^\dagger\phi_2 + \phi_3^\dagger\phi_3\right] +
\lambda \left[\phi_1^\dagger\phi_1 + \phi_2^\dagger\phi_2 + \phi_3^\dagger\phi_3\right]^2  \label{VTh}\\
&&\hspace{-8mm} + \lambda' \left[(\phi_1^\dagger\phi_1)(\phi_2^\dagger\phi_2) + (\phi_2^\dagger\phi_2)(\phi_3^\dagger\phi_3) + 
(\phi_3^\dagger\phi_3)(\phi_1^\dagger\phi_1)\right]
+ \lambda'' \left(|\phi_1^\dagger\phi_2|^2 + |\phi_2^\dagger\phi_3|^2 + |\phi_3^\dagger\phi_1|^2\right)\nonumber\\
&& + \left(\tilde\lambda  \left[(\phi_1^\dagger\phi_2)^2 + (\phi_2^\dagger\phi_3)^2 + (\phi_3^\dagger\phi_1)^2\right] + h.c.\right)\nonumber
\eea
with complex $\tilde\lambda$.
Its symmetry group is generated by independent sign flips of the individual doublets, by cyclic permutations of $\phi_1$, $\phi_2$, $\phi_3$,
and by the exchange of any pair of doublet together with the $CP$-transformation.
An alternative form of this potential is
\bea
V_{A_4 \rtimes \Z_2^*} &=& -m^2 \left[\phi_1^\dagger\phi_1 + \phi_2^\dagger\phi_2 + \phi_3^\dagger\phi_3\right] +
\lambda \left[\phi_1^\dagger\phi_1 + \phi_2^\dagger\phi_2 + \phi_3^\dagger\phi_3\right]^2  \label{VTh2}\\
&& + \lambda' \left[(\phi_1^\dagger\phi_1)(\phi_2^\dagger\phi_2) + (\phi_2^\dagger\phi_2)(\phi_3^\dagger\phi_3) + 
(\phi_3^\dagger\phi_3)(\phi_1^\dagger\phi_1)\right]\nonumber\\
&& \hspace{-1.2cm}+ \lambda_{\mathrm{Re}}\left[(\Re\phi_1^\dagger \phi_2)^2 + (\Re\phi_2^\dagger \phi_3)^2 + (\Re\phi_3^\dagger \phi_1)^2\right]
+ \lambda_{\mathrm{Im}}\left[(\Im\phi_1^\dagger \phi_2)^2 + (\Im\phi_2^\dagger \phi_3)^2 + (\Im\phi_3^\dagger \phi_1)^2\right]\nonumber\\
&& + \lambda_{\mathrm{ReIm}}\left[\Re\phi_1^\dagger \phi_2\, \Im\phi_1^\dagger \phi_2 + \Re\phi_2^\dagger \phi_3\,\Im\phi_2^\dagger \phi_3
+ \Re\phi_3^\dagger \phi_1\, \Im\phi_3^\dagger \phi_1\right]\,.\nonumber
\eea

{\bf \boldmath Group $S_4 \times \Z_2^*$.}
If the parameter $\tilde \lambda$ in (\ref{VTh}) is real or, equivalently, $ \lambda_{\mathrm{ReIm}} =0$ in (\ref{VTh2}), 
the potential becomes symmetric under $S_4 \times \Z_2^*$ generated by sign flips,
all permutation of the three doublets, and the $CP$-transformation.

{\bf \boldmath Group $(\Z_3\times\Z_3)\rtimes\Z_2 \simeq \Delta(54)/\Z_3$.} 
Consider the following potential 
\bea
V_{\Delta(54)/\Z_3} & = &  - m^2 \left[\phi_1^\dagger \phi_1+ \phi_2^\dagger \phi_2+\phi_3^\dagger \phi_3\right]
+ \lambda_0 \left[\phi_1^\dagger \phi_1+ \phi_2^\dagger \phi_2+\phi_3^\dagger \phi_3\right]^2 \nonumber\\
&&+ \lambda_1  \left[(\phi_1^\dagger \phi_1)^2+ (\phi_2^\dagger \phi_2)^2+(\phi_3^\dagger \phi_3)^2
- (\phi_1^\dagger \phi_1)(\phi_2^\dagger \phi_2) - (\phi_2^\dagger \phi_2)(\phi_3^\dagger \phi_3)
- (\phi_3^\dagger \phi_3)(\phi_1^\dagger \phi_1)\right]\nonumber\\
&&+ \lambda_2 \left[|\phi_1^\dagger \phi_2|^2 + |\phi_2^\dagger \phi_3|^2 + |\phi_3^\dagger \phi_1|^2\right] \nonumber\\
&&+ \lambda_3 \left[(\phi_1^\dagger \phi_2)(\phi_1^\dagger \phi_3) + (\phi_2^\dagger \phi_3)(\phi_2^\dagger \phi_1) + (\phi_3^\dagger 
\phi_1)(\phi_3^\dagger \phi_2)\right]
+ h.c.\label{VZ3Z3b}
\eea
with generic real $m^2$, $\lambda_0$, $\lambda_1$, $\lambda_2$ and complex $\lambda_3$. 
The symmetry group of this potential is $(\Z_3\times\Z_3)\rtimes\Z_2 = \Delta(54)/\Z_3$. Here, $\Delta(54)$ is generated by
the same $a_3$ and $b$ as before and, in addition, by the cyclic permutation 
\be
c = \left(\begin{array}{ccc} 0 & 1 & 0 \\ 0 & 0 & 1 \\ 1 & 0 & 0 \end{array}\right)\,,
\ee
while the subgroup $\Z_3$ is the center of $SU(3)$.

{\bf \boldmath Group $(\Z_3\times\Z_3)\rtimes(\Z_2\times \Z_2^*)$.}
The potential (\ref{VZ3Z3b}) becomes symmetric under a generalized-$CP$ transformation
if $\lambda_3 = k\cdot \pi/3$ with any integer $k$. In this case, one can make $\lambda_3$ real by a rephasing transformation.
The extra generator then is the $CP$-transformation.

{\bf \boldmath Group $\Sigma(36)\rtimes \Z_2^*$.}
The same potential (\ref{VZ3Z3b}) becomes symmetric under the group $\Sigma(36)\rtimes \Z_2^*$
if, upon rephasing, $\lambda_3 = (3\lambda_1 - \lambda_2)/2$.
The potential can then be rewritten as
\bea
V_{\Sigma(36)\rtimes \Z_2^*} & = & - m^2 I_0 + \lambda_0 I_0^2 + 3 \lambda_1 I_1\nonumber\\
&+& {\lambda_2 - 3\lambda_1 \over 2}
\left(|\phi_1^\dagger \phi_2 - \phi_2^\dagger \phi_3|^2 + |\phi_2^\dagger \phi_3 - \phi_3^\dagger \phi_1|^2 + |\phi_3^\dagger \phi_1 - 
\phi_1^\dagger \phi_2|^2\right)\,.
\label{VZ3Z3D8b}
\eea
Here $I_0$ and $I_1$ are the $SU(3)$-invariants
\bea
I_0 &=& {r_0 \over \sqrt{3}} = \phi_1^\dagger \phi_1+ \phi_2^\dagger \phi_2+\phi_3^\dagger \phi_3\,,\nonumber\\
I_1 & = & \sum_i r_i^2 = {(\phi_1^\dagger \phi_1)^2+ (\phi_2^\dagger \phi_2)^2+(\phi_3^\dagger \phi_3)^2
- (\phi_1^\dagger \phi_1)(\phi_2^\dagger \phi_2) - (\phi_2^\dagger \phi_2)(\phi_3^\dagger \phi_3)
- (\phi_3^\dagger \phi_3)(\phi_1^\dagger \phi_1)\over 3} \nonumber\\
&&\qquad+ |\phi_1^\dagger \phi_2|^2 + |\phi_2^\dagger \phi_3|^2 + |\phi_3^\dagger \phi_1|^2\,.
\eea
It is remarkable that this potential has only one ``structural'' free parameter, 
and the term containing it reduces the full $SU(3)$ symmetry group to a finite subgroup $\Sigma(36)$.


\end{document}